\documentclass[aps,prc,reprint,showpacs,groupedaddress,onecolumn]{revtex4-1}

\usepackage{graphicx,color} 
\usepackage{dcolumn}  
\usepackage{bm}       
\usepackage{amsmath}
\usepackage{enumerate}

\begin{document}

\newcommand {\nc} {\newcommand}
\newcommand{\vv}[1]{{$\bf {#1}$}}
\newcommand{\ul}[1]{\underline{#1}}
\def\bsigma{\mbox{\boldmath$\sigma$}}

\nc {\IR} [1]{\textcolor{red}{#1}}
\nc {\IB} [1]{\textcolor{blue}{#1}}
\nc {\IP} [1]{\textcolor{magenta}{#1}}

\title{Separable Representation of Multichannel Nucleon-Nucleus Optical Potentials} 

\author{L.~Hlophe$^{(a)}$}
\email{hlophe@nscl.msu.edu}
\author{Ch.~Elster$^{(b)}$}
\email{elster@ohio.edu}

\affiliation{
(a) National Superconducting Cyclotron Laboratory and Department of Physics and Astronomy, Michigan State University, East Lansing, MI 48824, USA \\
(b)Institute of Nuclear and Particle Physics,  and
Department of Physics and Astronomy,  Ohio University, Athens, OH 45701 \\
}

\date{\today}

\begin{abstract}
\begin{description}
\item[Background]
   One important ingredient for many applications of nuclear physics to astrophysics,
  nuclear energy,  and stockpile stewardship are cross sections for reactions of
neutrons with rare isotopes.
  Since direct measurements are often not feasible, indirect methods, e.g. (d,p)
reactions, should be used.
  Those (d,p) reactions may be viewed as three-body reactions and described with
Faddeev techniques.

\item[Purpose] Faddeev equations in momentum space have a long tradition of
  utilizing separable interactions in order to arrive at sets of coupled integral
  equations in one variable. Optical potentials representing the effective
interactions in the neutron (proton) nucleus subsystem are usually non-Hermitian as well as
energy-dependent. Including excitations of the nucleus in the calculation requires a multichannel
optical potential.  The purpose of this paper
is to introduce a separable, energy-dependent multichannel representation of complex,
energy-dependent optical potentials that contain excitations of the nucleus and
 that fulfill reciprocity exactly.

\item[Methods] Momentum space Lippmann-Schwinger integral equations are solved with
standard techniques to
obtain the form factors for the separable representation.

\item[Results] Starting from energy-dependent multichannel optical potentials for neutron and proton
scattering from $^{12}$C, separable representations based on a 
generalization of the Ernst-Shakin-Thaler (EST) scheme are constructed which fulfill
reciprocity exactly. Applications to
n$+^{12}$C and  p$+^{12}$C scattering are investigated for energies from 0 to 50~MeV. 

\item[Conclusions] We find that the energy-dependent separable representation
of complex, energy-dependent phenomenological multichannel optical potentials
describes scattering data with the same quality as the original potential.
\end{description}
\end{abstract}


\maketitle

\section{Introduction}
\label{intro}

Nuclear reactions are an important probe to learn about the structure of unstable nuclei.
Due to the short lifetimes
involved, direct measurements are usually not possible. Therefore indirect measurements
 using ($d,p$) reactions have been proposed (see e.g.
  Refs.~\cite{RevModPhys.84.353,jolie,Kozub:2012ka}).
  Deuteron induced reactions are particularly attractive from an experimental
perspective,
  since deuterated targets are readily available. From a theoretical perspective they are
  equally attractive because the scattering problem can be reduced to an effective
  three-body problem~\cite{Nunes:2011cv}. Traditionally deuteron-induced single-neutron transfer
($d,p$) reactions have been used to study the shell structure in stable nuclei. Nowadays
  experimental techniques are available to apply the same approaches to exotic beams (see
  e.g.~\cite{Schmitt:2012bt}).
  Deuteron induced $(d,p)$ or $(d,n)$ reactions in inverse kinematics are
  also useful to extract neutron or proton capture rates on unstable nuclei of
  astrophysical
  relevance.~\cite{Kankainen:2016zcf}. Given the many ongoing experimental programs  worldwide using these
reactions, a reliable reaction theory for $(d,p)$ reactions is critical.

One of the most challenging aspects of solving the three-body problem for nuclear
reactions is the repulsive Coulomb interaction. While for very light
  nuclei, exact calculations of (d,p) reactions based on momentum-space Faddeev
equations in
  the Alt-Grassberger-Sandhas (AGS)~\cite{ags} formulation can be carried
  out~\cite{Deltuva:2009fp} by using a screening and renormalization
  procedure~~\cite{Deltuva:2005wx,Deltuva:2005cc}, this technique leads to increasing
technical
  difficulties when moving to computing (d,p) reactions with
  heavier nuclei~\cite{hites-proc}. Therefore, a new formulation of the Faddeev-AGS
equations, which does not rely on a screening procedure, was presented in
  Ref.~\cite{Mukhamedzhanov:2012qv}.
Here the Faddeev-AGS equations are cast in a
  momentum-space  Coulomb-distorted partial-wave representation
  instead of the plane-wave basis.  Thus all operators,  specifically the interactions
  in the two-body subsystems, must be evaluated in the Coulomb basis, which is a
  nontrivial task.
  The formulation of Ref.~\cite{Mukhamedzhanov:2012qv} requires the interactions   in the
  subsystems to be of separable form.
The same reference suggests an extension of the Faddeev-AGS equations to take excitations
of the nucleus into account. Faddeev-AGS calculations taking into account rotational
excitations were carried out for (d,p) reactions with $^{10}$Be and $^{24}$Mg in
Refs.\cite{Deltuva:2013jna,Deltuva:2016muy}, showing that the inclusion of excited states
in those nuclei could improve the description of experimental data. 

To include excitations of the nucleus in the formulation of
Ref.~\cite{Mukhamedzhanov:2012qv}, separable representations of the effective neutron and
proton interactions with the nucleus which include those excitations need to be constructed.
We follow here the work already carried out for single-channel optical potentials that are
complex~\cite{Hlophe:2013xca}  as well as energy dependent~\cite{Hlophe:2016}.
These separable representations have roots in the work by Ernst, Shakin, and
Thaler~\cite{Ernst:1973zzb} (EST), who derived a scheme for constructing separable
representations using scattering wave functions at given energies as basis for their
expansion. In momentum space, the scattering wave functions are related to half-shell $t$
matrices. Thus, the EST choice guarantees that, at those fixed energies, the separable
expansion is not only on-shell but also half-shell exact. 
The generalization to complex, energy-dependent optical potentials of
Refs.~\cite{Hlophe:2013xca,Hlophe:2016} fulfills the same conditions
and in addition guarantees that 
reciprocity is fulfilled.

Taking into account excitations  of the nucleus leads to multichannel optical
potentials. Although an extension of the EST scheme to multichannel potentials was
already carried out in Ref.~\cite{Pieper74}, that work
was limited to Hermitian potentials. In this manuscript we extend the work of
Refs.~\cite{Hlophe:2013xca,Hlophe:2016} to obtain a separable representation
of complex, energy-dependent multichannel optical potentials.  First we derive the
formulation for multichannel neutron-nucleus optical potentials in Sec.~\ref{mch-nopt} 
and apply the formulation to neutron elastic and inelastic scattering off $^{12}$C. 
In Sec.~\ref{mch-popt} we extend this formulation to charged particle scattering and
show as application the elastic and inelastic scattering of protons from $^{12}$C.
Our findings are summarized in Sec.~\ref{summary}. The manuscript is accompanied by three
appendices explaining analytic and numerical details.


\section{Energy-Dependent Multichannel Neutron-Nucleus Optical Potentials}
\label{mch-nopt}

\subsection{Formal Considerations}
\label{nopt-form}

\subsubsection{Coupled-Channel Formalism }

For setting up the multichannel problem, let us consider a nucleus
characterized by the spin-parity $I^\pi$. The
nuclear wavefunction is represented by $|\Phi_{IM_I}\rangle$, where
$M_I$ is the projection of $I$ along the $z$-axis.  
A neutron interacting with the
nucleus through a potential $U$ has an angular momentum $j_p=l\pm s$, with $l$
being the relative orbital angular momentum and  $s=1/2$  its spin. The total, conserved
angular momentum ${\bf J}$ of the system takes values  $|I-j_p|\le J\equiv |{\bf J}|\le
I+j_p$, and its projection along the z-axis is given by $M$. 
The states of conserved angular momentum are thus given by
  \begin{eqnarray}
  |(Ilsj_p)JM\rangle = \sum\limits_{M_Im_{j_p}} C(Ij_pJ,M_Im_{j_p}M)|{\cal Y}_{ls}^{j_pm_{j_p}}\rangle|\Phi_{IM_{I}}\rangle,
 \label{eq:npot-1}
 \end{eqnarray}
where
 \begin{equation}
  |\mathcal{Y}_{ls}^{j_pm_{j_p}}\rangle= \sum\limits_{m_lm_s} C(lsj_p,m_lm_sm_{j_p})|Y_{lm_l}\rangle|sm_s\rangle.
  \label{eq:nad0a-1}
 \end{equation}
 The functions $Y_{lm_l}$  are the spherical harmonics, $\chi_{sm_s}$ are the corresponding
spinors, and $C(lsj_p,m_lm_sm_{j_p})$ Clebsch-Gordon (C.G.) coefficients. 
The quantities $I$, $l$, and $j_p$  collectively characterize a particular
configuration of the system but are not individually conserved.
 These configurations make up the angular momentum channels and will be denoted using the
Greek letters $\alpha,\;\beta,\;\gamma,$ etc. Since the potential $U$ couples different
angular momentum channels, we need to employ 
a coupled-channel formulation to describe a scattering process. The multichannel wavefunction $|\Psi_{\alpha_0}^{J\pi(+)}\rangle$ is characterized
 by the total angular momentum $J$, the parity $\pi$, as well as the incident angular 
 momentum channel $\alpha_0$, and obeys
 a coupled set of Lippmann-Schwinger (LS)
 equations,
  \begin{equation}
   |\Psi^{J(+)}_{\alpha\alpha_0}\rangle=|lk_0\rangle\;\delta_{\alpha\alpha_0} + G_{0\alpha}(E)\sum\limits_{\alpha'} 
   U_{\alpha\alpha'}^J|\Psi^{J(+)}_{\alpha'\alpha_0}\rangle.
  \label{eq:mch2g}
 \end{equation}
  Here $|\Psi^{J(+)}_{\alpha\alpha_0}\rangle$ represents a projection
  of the wavefunction $|\Psi_{\alpha_0}^{J\pi(+)}\rangle$ onto the 
  $\alpha$-channel.     
  The free propagator in a given channel $\alpha$ has the form 
$G_{0\alpha}(E)=\left(E-\varepsilon_{\alpha}-H_0+i\epsilon\right)^{-1}$  with $E=k_0^2/2\mu$
being the nonrelativistic kinetic energy in the incident channel and $\mu$ the reduces mass.
 Here $\varepsilon_\alpha$
 designates the nuclear excitation energy in channel $\alpha$.  
Using the states of Eq.~(\ref{eq:mch2g}) in  
the operator LS equation, $T(E)=U+UG_0(E)T(E)$, one obtains
a set of coupled $t$ matrix equations
 \begin{equation}
  T_{\alpha\alpha_0}^J(E)=U_{\alpha\alpha_0}^J+
  \sum\limits_{\alpha'}U_{\alpha\alpha'}^JG_{0\alpha'}(E)
  T_{\alpha'\alpha_0}^J(E).
  \label{eq:multi2g}
  \end{equation}
Here $U_{\alpha\alpha_0}^J\equiv\langle\alpha JM|U|\alpha_0 JM\rangle$
and $T_{\alpha\alpha_0}^J(E)\equiv\langle\alpha JM|T(E)|\alpha_0 JM\rangle$ are elements of the
 multichannel potential and $t$ matrix respectively. The explicit LS equation for the half-shell
 $t$ matrix in momentum space takes the form
 \begin{equation}
  T_{\alpha\alpha_0}^J(k',k;E)=U_{\alpha\alpha_0}^J(k',k)
  +\sum\limits_{\alpha'}\int\limits_{0}^\infty dp p^2\;U_{\alpha\alpha'}^J(k,p)G_{0\alpha'}(E,p)
  T_{\alpha'\alpha_0}^J(p,k;E),
  \label{eq:multi3g}
  \end{equation}
 where the propagator is given by $G_{0\alpha}(E,p)=\left(E_\alpha-p^2/2\mu_{\alpha}+i\epsilon\right)^{-1}$. Here $E_\alpha\equiv E-\varepsilon_{\alpha}$
 is the center of mass (c.m.) energy  and $\mu_\alpha$ the reduced mass in channel $\alpha$.  Further details on the
 scattering of neutrons from deformed nuclei are given in Appendix~\ref{appx-B}.

\subsubsection{Separable Representation of complex, energy-dependent Multichannel Potentials}
\label{nopt-eest} 

For  single-channel separable representations,  
the work of Ref.~\cite{Hlophe:2013xca}
already showed that the EST scheme presented in Ref.~\cite{Ernst:1973zzb}
leads to the violation of reciprocity when applied to complex
potentials, which can be resolved if  
both, incoming and outgoing scattering states are used in the separable expansion. 
However, the separable potentials from Ref.~\cite{Hlophe:2013xca} still do not completely fulfill reciprocity for
energy-dependent potentials.
A further generalization of the EST scheme for potentials that are both
complex and energy-dependent had to be developed~\cite{Hlophe:2016}, which
 involves  introducing an energy-dependent separable potential
\begin{eqnarray}
u(E) = \sum\limits_{ij}U(E_i)\big|\psi_i^{(+)}\big\rangle\lambda_{ij}(E)\big\langle\psi_j^{(-)}\big|U(E_j).
\label{eq:form9}
\end{eqnarray}
Here $|\psi_i^{(+)}\rangle$ is an outgoing 
single-channel scattering wavefunction corresponding to $U$
at the energy $E_i$, and $|\psi_i^{(-)}\rangle$ is an incoming 
single-channel scattering wavefunction corresponding to $U^*$
at the energy $E_i$. The energies $E_i$ are referred to as the  
EST support points. As Eq.~(\ref{eq:form9}) shows, the energy dependency
is introduced into the coupling matrix elements $\lambda_{ij}(E)$, which
obey the constraint
\begin{eqnarray}
{\cal U}^e_{mn}(E)&=& 
  \langle\psi_m^{(-)}|u(E)\big|\psi_n^{(+)}\rangle\cr\cr
  &=&\sum\limits_{ij}\langle\psi_m^{(-)}\big|U(E_i)\big|\psi_i^+\big\rangle\lambda_{ij}(E)
  \big\langle\psi_j^{(-)}\big|U(E_j)\big|\psi_n^+\big\rangle,
  \label{eq:form10}
 \end{eqnarray}
where the matrix elements between in- and out-going scattering states
at energies $E_m$ and $E_n$ are given as
\begin{eqnarray}
{\cal U}^e_{mn}(E)&\equiv& \big\langle\psi_m^{(-)}\big|U(E)\big|\psi_n^{(+)}\big\rangle.
\label{eq:from10a}
\end{eqnarray}
Defining the matrix elements ${\cal U}_{mn}$
\begin{equation}
\mathcal{U}_{mn} \equiv{\cal U}^e(E_m)=\langle\psi_m^{(-)}|U(E_m)|\psi_n^{(+)}\rangle,
\label{eq:form10b}
\end{equation}
leads to a more compact form for Eq.~(\ref{eq:form10}),
 \begin{eqnarray}
{\cal U}^e_{mn}(E) &=&\sum\limits_{ij}{\cal U}_{mi}^t\;\lambda_{ij}(E)\;{\cal U}_{jn},\cr
                   &=&\left[{\cal U}^t\cdot\lambda(E)\cdot {\cal U} \right]_{mn} 
   \label{eq:form10d}
\end{eqnarray}
where ${\cal U}^t$ represents the transpose of ${\cal U}$. 
This constraint ensures that the eigenstates of $U$ and
$u$ coincide at the EST support points. The coupling matrix
is symmetric so that $\lambda_{ij}(E)=\lambda_{ji}(E)$, a condition
necessary for the fulfillment of reciprocity.
Since $u$ is explicitly energy dependent, we coin this representation as
energy-dependent EST (eEST) scheme.  For single-channel, energy-dependent optical
potentials the on-shell $t$ matrix elements obtained with the eEST scheme and
its energy-independent approximation (EST) agree quite well~\cite{Hlophe:2016}.

To generalize the eEST scheme to multichannel potentials, we proceed analogously to Ref.~\cite{Pieper74} and replace the single-channel
scattering wavefunctions with their multichannel counterparts, leading to a 
multichannel separable potential      
 \begin{eqnarray}
  u(E)&=&\sum\limits_{\rho\sigma}\sum\limits_{ij} 
  \left(\sum\limits_{\gamma JM} U(E_i)\big|\gamma JM\; \Psi_{\gamma\rho,i}^{J(+)}\big\rangle\right) \;
  \lambda_{ij} ^{\rho\sigma}(E)\;
 \left(\sum\limits_{\gamma JM}\big\langle \Psi_{\gamma\sigma,j}^{J(-)}\;\gamma JM\big|U(E_j)\right).
\label{eq:mch1a-1}
 \end{eqnarray}
The indices $i$ and $j$ stand for the EST support points.  Using
the definition of the multichannel half-shell $t$ matrix~\cite{Gloecklebook},
 \begin{eqnarray}
  T(E_i)\big|\rho JM\;k_i^\rho\big\rangle=\sum\limits_{\gamma} U(E_i)\big|\gamma\;JM\Psi_{{\gamma\rho}}^{J(+)}\big\rangle,
  \label{eq:nad3c}
  \end{eqnarray}
Eq.~(\ref{eq:mch1a-1}) can be recast as
\begin{eqnarray}
  u(E)&=&\sum\limits_{JM}\sum\limits_{J'M'}\sum\limits_{\rho\sigma}\sum\limits_{ij} 
   T(E_i)\big|\rho JM\; k_i^\rho\big\rangle \;\lambda_{ij}
  ^{\rho\sigma}(E)\;
 \big\langle k_j^\sigma\;\sigma J'M'\big|T(E_j).
 \label{eq:mch21a}
\end{eqnarray}
To determine the constraint on $u(E)$, 
 we first generalize 
 the matrices ${\cal U}^{e}(E)$ and ${\cal U}$
 defined by Eqs.~(\ref{eq:from10a}) and~(\ref{eq:form10b}) to
 multichannel potentials. This is accomplished by replacing
 the single-channel scattering states by the multichannel
 wavefunctions so that 
 \begin{eqnarray}
  {\cal U}^{e,\alpha\beta}_{mn}(E)&\equiv&\sum\limits_{\gamma\nu}\big\langle \Psi_{\gamma\alpha,m}^{J(-)}\;\gamma JM\big|
  U(E)\big|\nu JM\;\Psi_{\nu\beta,n}^{J(+)}\rangle,\cr\cr
   &=&\sum\limits_{\gamma\nu}\big\langle \Psi_{\gamma\alpha,m}^{J(-)}\big|
   U_{\gamma\nu}^J(E)\big|\Psi_{\nu\beta,n}^{J(+)}\big\rangle,
   \label{eq:mch21b0}
 \end{eqnarray}
and 
\begin{eqnarray}
  {\cal U}^{\alpha\beta}_{mn}\equiv{\cal U}^{e,\alpha\beta}_{mn}(E_m)&
   =\sum\limits_{\gamma\nu}\big\langle \Psi_{\gamma\alpha,m}^{J(-)}\big|
   U_{\gamma\nu}^J(E_m)\big|\Psi_{\nu\beta,n}^{J(+)}\big\rangle.
   \label{eq:mch21b1}
 \end{eqnarray}

The $J$ dependence of matrix elements ${\cal U}^{e,\alpha\beta}_{mn}(E)$
and ${\cal U}^{\alpha\beta}_{mn}$ is omitted for simplicity.
One one hand, Eq.~(\ref{eq:mch21b1}) shows that the matrix 
${\cal U}$ depends only on
the support energies $E_m$ and $E_n$. On other hand, we see
from Eq.~(\ref{eq:mch21b0}) that ${\cal U}^{e}(E)$ depends
on the projectile energy $E$ as well as the support energies. 
The  constraint on the separable potential is obtained
by substituting the multichannel matrices ${\cal U}^e$ and ${\cal U}$ into
Eq.~(\ref{eq:form10d}) leading to    
 \begin{eqnarray}
{\cal U}^{e,\alpha\beta}_{mn}(E)
 &=& \sum\limits_{\rho\sigma}\sum\limits_{ij}\big({\cal U}^t\big)^{\alpha\rho}_{mi}\;\lambda_{ij}^{\rho\sigma}(E)\;{\cal U}^{\sigma\beta}_{jn},\cr
&=&\left[{\cal U}^t\cdot\lambda(E)\cdot {\cal U} \right]_{mn}^{\alpha\beta}. 
  \label{eq:mch21b2}
 \end{eqnarray}

To evaluate the separable multichannel $t$ matrix, we 
insert Eqs.~(\ref{eq:mch21a})-(\ref{eq:mch21b2}) into
Eq.~(\ref{eq:multi2g}) and obtain
 \begin{eqnarray}
  t(E)&=&\sum\limits_{\rho\sigma}\sum\limits_{ij} 
  \left(\sum\limits_{\gamma JM} U(E_i)\big|\gamma JM\; \Psi_{\gamma\rho,i}^{J(+)}\big\rangle\right) \;\tau_{ij}
  ^{\rho\sigma}(E)\;
 \left(\sum\limits_{\gamma JM}\big\langle \Psi_{\gamma\sigma,j}^{J(-)}\;\gamma JM\big|U(E_j)\right),\cr\cr
 &=&\sum\limits_{JM}\sum\limits_{J'M'}\sum\limits_{\rho\sigma}\sum\limits_{ij} 
   T(E_i)\big|\rho JM\; k_i^\rho\big\rangle \;\tau_{ij}
  ^{\rho\sigma}(E)\;
 \big\langle k_j^\sigma\;\sigma J'M'\big|T(E_j).
 \label{eq:mch21c}
 \end{eqnarray}
The coupling matrix elements $\tau_{ij}^{\rho\sigma}(E)$ fulfill
 \begin{eqnarray}
   R(E)\cdot\tau(E)=\mathcal{M}(E),
   \label{eq:mch24a}
 \end{eqnarray}   
 where
 \begin{eqnarray}
  R^{\rho\sigma}_{ij}(E)&=&\Big\langle k_i^{\rho}\Big|~T_{{\rho\sigma}}^J(E_i)~+~\sum\limits_{{\beta}}
  T_{{\rho\beta}}^J(E_i)G_{{\beta}}(E_j)T_{ {\beta\sigma}}^J(E_j)\Big| k_j^{\sigma}
  \Big\rangle\cr\cr
 &-&~\sum\limits_{ {\beta\beta}'}\sum\limits_{n}{\cal M}_{in}^{ {\rho\beta}}\langle  k_n^{\beta}\Big|
  T_{ {\beta\beta'}}^J(E_n)G_{ {\beta'}}(E)T_{ {\beta'\sigma}}^J(E_j)
 \Big| k_j^{\sigma}\Big\rangle,
   \label{eq:mch24b}
 \end{eqnarray} 
 and
 \begin{equation}
 {\cal M}^{\rho\sigma}_{ij}(E)=\left[{\cal U}^e(E)\cdot{\cal U}^{-1}\right]^{\rho\sigma}_{ij}.
 \label{eq:mch25c}
 \end{equation}
 The expression for the matrix $R^{\rho\sigma}_{ij}(E)$ is analogous to
 the one obtained in Ref.~\cite{Hlophe:2016} for the single-channel case
 except for the extra channel indices. The numerical implementation of Eqs.~(\ref{eq:mch21c})-(\ref{eq:mch25c}) is discussed in Appendix~\ref{appx-B}.  
 
The eEST scheme simplifies considerably for energy-independent potentials.
From Eqs.~(\ref{eq:mch21b0}) we see that the dependence of the separable potential
$u$ on the scattering energy $E$ arises from the energy dependence of $U$.
Consequently, the separable potential is independent of the energy $E$ if
$U$ is energy-independent. To derive the multichannel EST separable 
representation for energy-independent potentials we set
\begin{eqnarray}
 U(E)=U(E_i)=U(E_j)=U(E_m)=U
\label{eq:mch25f}
\end{eqnarray}
 in Eq.~(\ref{eq:mch21b1}). This leads to 
\begin{eqnarray}
  {\cal U}^{\alpha\beta}_{mn}\equiv{\cal U}^{e,\alpha\beta}_{mn}&
   =\sum\limits_{\gamma\nu}\big\langle \Psi_{\gamma\alpha,m}^{J(-)}\big|
   U_{\gamma\nu}^J\big|\Psi_{\nu\beta,n}^{J(+)}\big\rangle,
   \label{eq:mch25d}
 \end{eqnarray}
which implies that the matrices ${\cal U}$ and ${\cal U}^e$ are identical.
 As a consequence, the matrix ${\cal M}$ in Eq.~(\ref{eq:mch25c})
 reduces to an identity matrix.
The energy-independent separable potential takes the form
\begin{eqnarray}
  u&=&\sum\limits_{JM}\sum\limits_{J'M'}\sum\limits_{\rho\sigma}\sum\limits_{ij} 
   T(E_i)\big|\rho JM\; k_i^\rho\big\rangle \;\lambda_{ij}
  ^{\rho\sigma}\;
 \big\langle k_j^\sigma\;\sigma J'M'\big|T(E_j),
 \label{eq:mch21a-2}
\end{eqnarray}
with the constraint
 \begin{eqnarray}
\delta_{\alpha\beta}\delta_{mn}
 &=& \sum\limits_{\rho}\sum\limits_{i}\big({\cal U}^t\big)^{\alpha\rho}_{mi}\;\lambda^{\rho\beta}_{in}. 
  \label{eq:mch21b2-2}
 \end{eqnarray}
 The corresponding separable $t$ matrix is given by Eq.~(\ref{eq:mch21c}).
 The coupling matrix $\tau(E)$ is obtained by replacing the matrix ${\cal M}$
 in Eqs.~(\ref{eq:mch24a}) and (\ref{eq:mch24b}) with the identity matrix so that
 \begin{eqnarray}
   R(E)\cdot\tau(E)={\bf 1}.
   \label{eq:mch24a-1}
 \end{eqnarray}   
 The matrix elements $ R^{\rho\sigma}_{ij}(E)$ are given as
 \begin{eqnarray}
  R^{\rho\sigma}_{ij}(E)&=&\Big\langle k_i^{\rho}\Big|~T_{{\rho\sigma}}^J(E_i)~+~\sum\limits_{{\beta}}
  T_{{\rho\beta}}^J(E_i)G_{{\beta}}(E_j)T_{ {\beta\sigma}}^J(E_j)\Big| k_j^{\sigma}
  \Big\rangle\cr\cr
 &-&~\sum\limits_{ {\beta}}\sum\limits_{n}\langle  k_n^{\rho}\Big|
  T_{ {\rho\beta}}^J(E_n)G_{ {\beta}}(E)T_{ {\beta\sigma}}^J(E_j)
 \Big| k_j^{\sigma}\Big\rangle.
   \label{eq:mch24b-2}
 \end{eqnarray} 
Since the separable potential does not depend on the scattering energy $E$,
we refer to this scheme as the energy-independent EST separable representation. 
Although it is derived for energy-independent potentials, it can be applied to energy-dependent ones as well since only the multichannel half-shell $t$ matrices are required as input. The consequences of applying the energy-independent EST
separable representation scheme to energy-dependent potentials
were investigated in Ref.~\cite{Hlophe:2016} for single-channel optical potentials. It was determined that the off-shell $t$ matrix was not symmetric and thus violated the reciprocity theorem. Here a similar study
will be carried out for multichannel neutron
optical potentials. 
     
\subsection{ Elastic and Inelastic Scattering of Neutrons from $^{12}$C}  
\label{eest-nc12}

To illustrate the implementation of the multichannel eEST separable
representation scheme, we consider the scattering of neutrons from the
nucleus $^{12}$C. The $^{12}$C nucleus possesses  selected excited
states, with the first and second  levels having
 $I^\pi=2^+$  and $I^\pi=4^+$ and being  located at 4.43 and 14.08~MeV above the $0^+$ ground state. The collective rotational model~\cite{ThompsonNunes} is
 assumed to the coupling between the ground state and these
 excited states. 
 We consider here elastic scattering
 and inelastic scattering to the $2^+$ rotational state. 
To test the multichannel eEST separable representation we use the deformed optical potential model (DOMP) 
derived by Olsson et {\it al.}~\cite{Olsson:1989npa} and fitted to elastic and inelastic scattering data 
between 16 and 22~MeV laboratory kinetic energy.  DOMPs are generally  constructed by introducing angular
dependence to the optical model. Some details are provided in Appendix~\ref{domps}.

\subsubsection{$S$-matrix Elements and Differential Cross Section}
\begin{figure}
 \includegraphics[width=15cm]{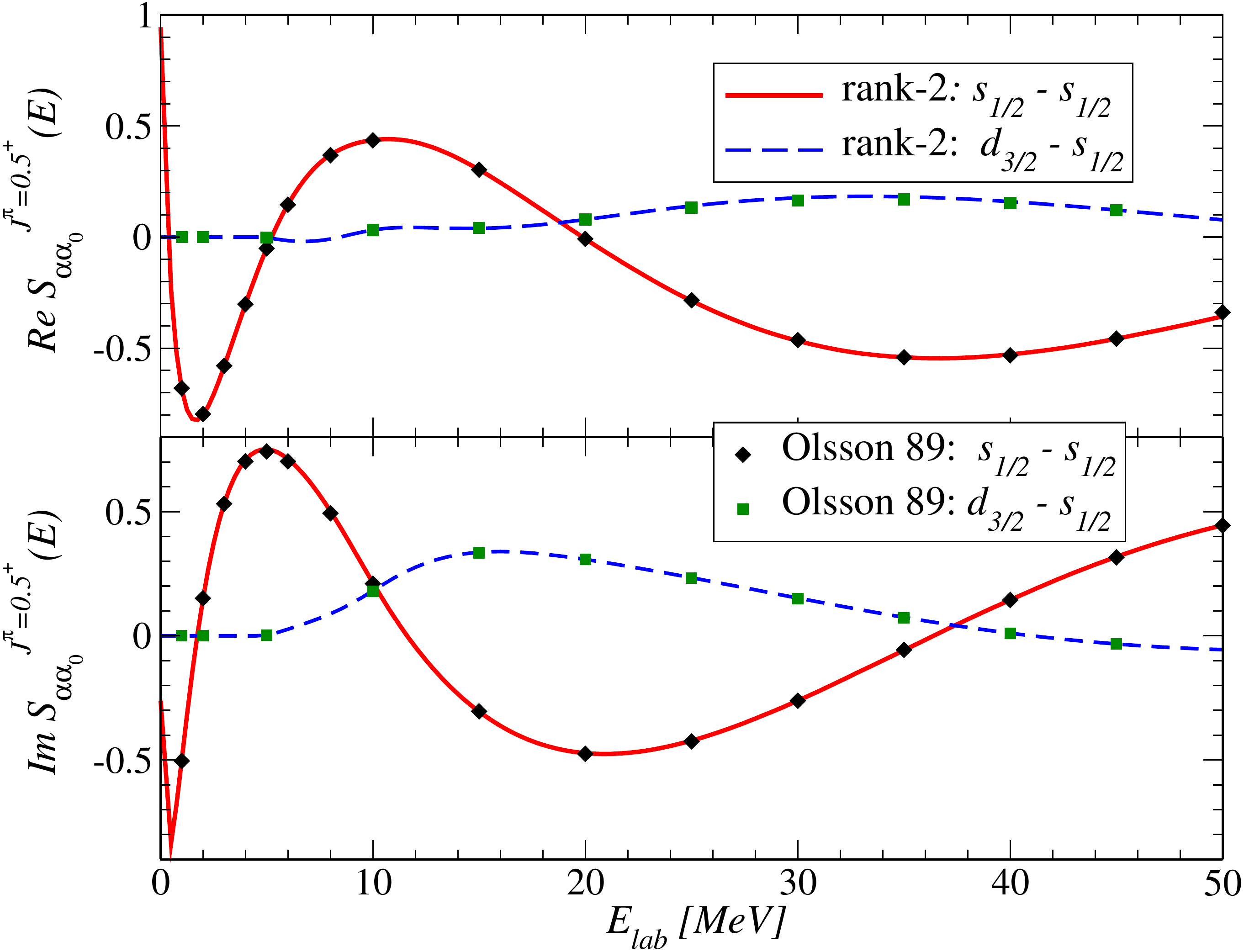}
 \caption{ The  energy-dependent EST (eEST) separable
    representation of the $J^\pi=1/2^+$ multichannel
   $S$-matrix elements,  $ S_{\alpha\alpha_0}^J(E)$,
    for the $n$+$^{12}$C system as function of the laboratory energy.
 The solid (dashed) line represents the $S$-matrix elements obtained with
  a rank-2 eEST separable representation in the 
$0^+\otimes s_{1/2}\rightarrow 0^+\otimes s_{1/2}$ ($0^+\otimes s_{1/2}\rightarrow 2^+\otimes
d_{3/2}$) channel. 
  The support points are located at 6 and 40~MeV.    
   The filled diamonds and squares represent the corresponding
  $S$-matrix elements directly
  evaluated with the Olsson~89 DOMP~\cite{Olsson:1989npa}. 
  \label{fig:fig1}
  }
  \end{figure}

First we want to consider $S$-matrix elements in well defined channels to study how well
the energy-dependent multichannel eEST scheme can represent them and what rank is required
to do so. In Fig.~\ref{fig:fig1} the $J^\pi=1/2^+$  S-matrix elements, $S_{\alpha\alpha_0}^J(E)$,
are shown for the  $0^+\otimes s_{1/2}\rightarrow 0^+\otimes s_{1/2}$ and
$0^+\otimes s_{1/2}\rightarrow 2^+\otimes d_{3/2}$ 
channels for neutron scattering from $^{12}$C as
function of the laboratory energy. The diagonal $0^+\otimes s_{1/2}\rightarrow 0^+\otimes s_{1/2}$
channel is represented by the solid line and the coupling to the $2^+\otimes d_{3/2}$ by the dashed line.    
The corresponding calculations using directly the Olsson~89 DOMP are represented by the solid diamonds
and squares. The agreement between the energy-dependent separable representation and the original
calculation is excellent. It should be noted that in this case already a rank-2 representation with 
support points at 6 and 40~MeV is
sufficient to achieve this high quality agreement between 0 and 50~MeV laboratory energy.

\begin{figure}
 \includegraphics[width=15cm]{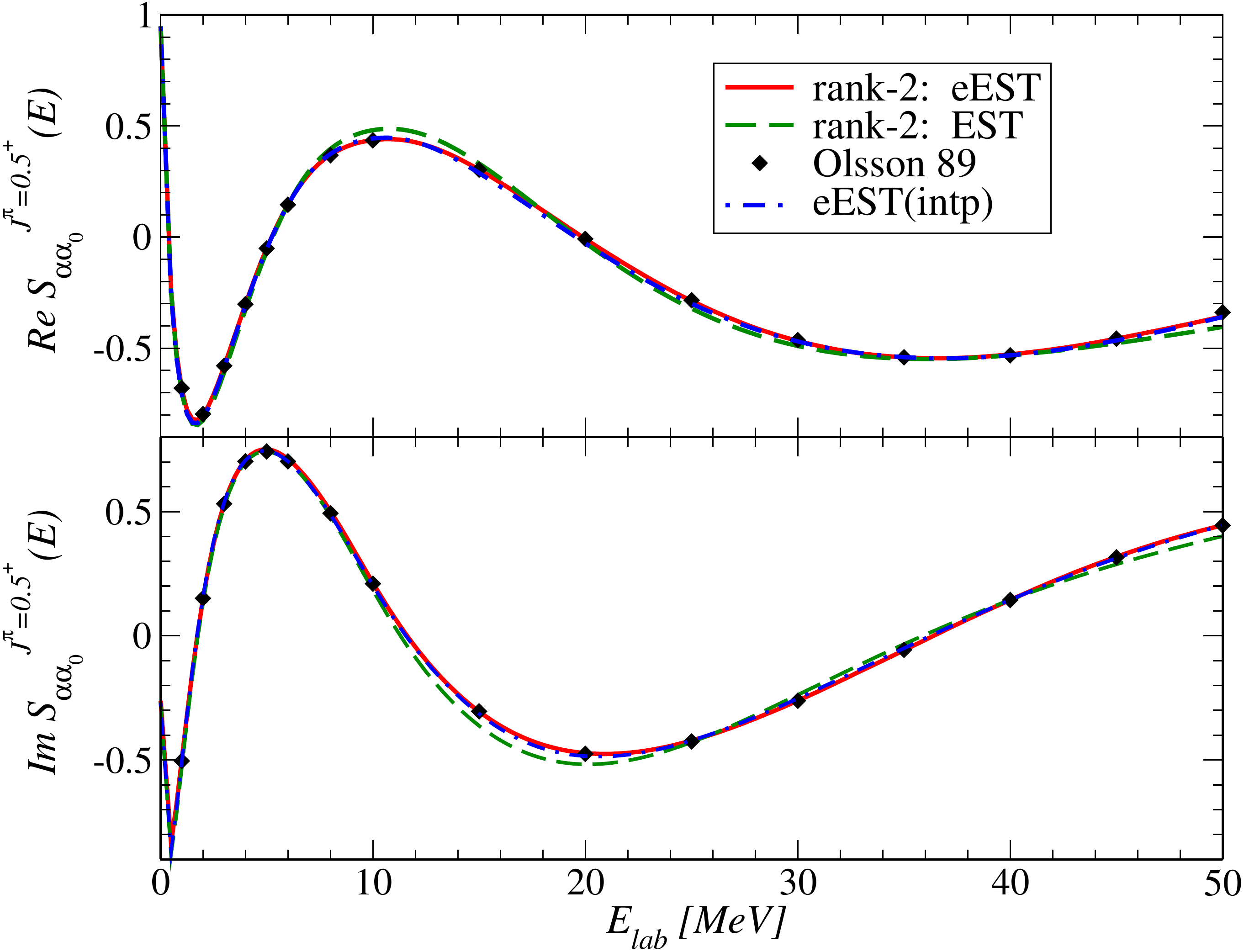}
 \caption{The $S$-matrix elements of the $0^+\otimes s_{1/2}\rightarrow 0^+\otimes s_{1/2}$ channel for
the $n$+$^{12}$C system as function of the laboratory energy. The solid line represent the rank-2
energy-dependent eEST separable representation, while the dashed line give the energy-independent EST
separable representation. For both the support points are at  6 and 40~MeV. The dash-dotted line shows
the $S$-matrix elements obtained with the interpolated eEST scheme. The filled diamonds represent the
calculation based directly on the Olsson~89 DOMP~\cite{Olsson:1989npa}.
  \label{fig:fig2}
  }  
 \end{figure}

The energy-independent EST scheme is a simplification of the eEST scheme leading to an energy-independent
separable representation. To illustrate the difference between the two schemes, Fig.~\ref{fig:fig2}
shows the $S$-matrix elements in the diagonal channel $0^+\otimes s_{1/2}\longrightarrow 0^+\otimes
s_{1/2}$ computed in the eEST scheme (solid line) and the energy-independent EST scheme (dashed line)
using the same support points (rank). As reference the calculation with the original Olsson~89
potential is given by the filled diamonds. The figure clearly shows that the representation with the
EST scheme is of lesser quality than the eEST scheme. This finding is consistent with observations for
representations of single-channel optical potentials~\cite{Hlophe:2013xca}. This suggests that 
the EST scheme might be improved by increasing the rank of the representation. However, in the
multichannel case there is an additional complication, since in the energy-independent scheme coupling matrix elements are not symmetric, i.e. $S^J_{\alpha\alpha'} \neq S^J_{\alpha'\alpha}$,
as will be illustrated later. 
  
\begin{figure}
 \includegraphics[width=15cm]{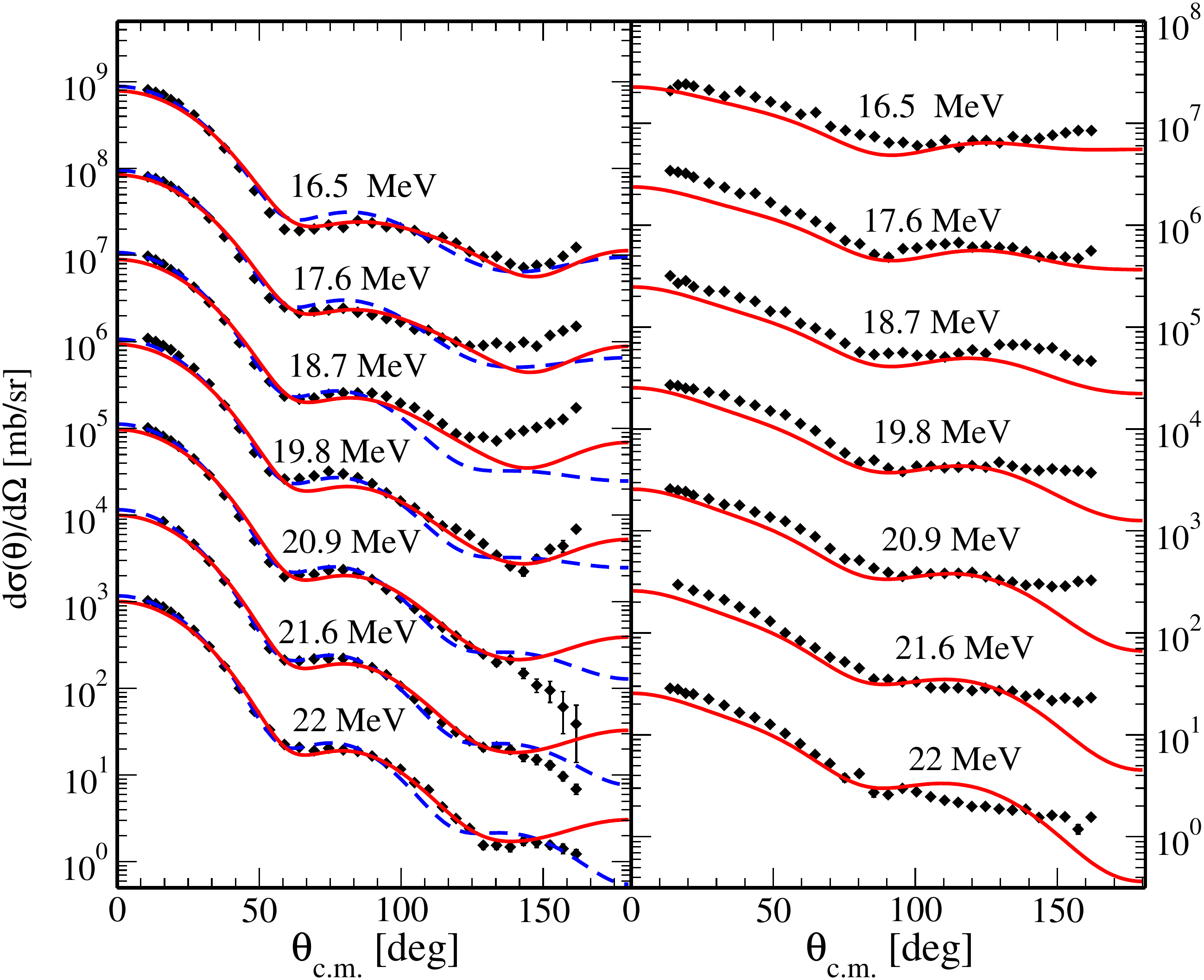}
 \caption{ The differential cross sections for scattering in the  $n+^{12}$C system
   computed at different incident neutron energies with the eEST separable representation of the
Olsson~89 DOMP~\cite{Olsson:1989npa} (solid lines). 
The left hand panel shows the differential cross section for elastic scattering, while  the right 
hand panel depicts the differential cross section for inelastic scattering to  
the $2^+$ state of $^{12}$C.  The dashed 
lines indicate cross sections computed with the spherical
Olsson~89~\cite{Olsson:1989npa} OMP. 
The filled diamonds represent  the data taken from Ref.~\cite{Olsson:1989npa}.  
The cross sections are scaled up by multiples of 10. 
The results at 21.6~MeV are multiplied by 10, those at 20.9 
are multiplied by 100, etc.       
  \label{fig:fig3}
  }
 \end{figure}

The numerical effort needed to evaluate the matrix elements
  ${\cal U}^{e,\alpha\beta}_{ij}(E)$ increases rapidly with
  the number of coupled channels. To simplify the implementation
  of the eEST scheme, one can avoid the evaluation
  of these matrix elements for every energy one wants to compute.
 Instead ${\cal U}^{e,\alpha\beta}_{ij}(E)$ 
   can be computed at fixed energies and an interpolation
   scheme used to determine its value elsewhere. Following
   Ref.~\cite{Hlophe:2016}, we choose
   the fixed energies to coincide with the EST support points. Such a choice
   has the advantage that the potential matrix elements are needed only
   at the support points. The results obtained with the interpolated eEST
   scheme are shown by the dash-dotted lines in Fig.~\ref{fig:fig2}. We
   observe that the results agree remarkably well with those obtained with
   the exact eEST scheme. 
To show the overall quality of the eEST separable representation in all partial wave $S$ matrix
elements, we compute cross sections for elastic and inelastic scattering.       
 In Fig.~\ref{fig:fig3} the differential
    cross sections for elastic and inelastic scattering for the $n+^{12}$C system are shown at various
    incident neutron energies. The left hand panel shows the 
     differential cross section for elastic scattering, and  the right hand panel the
 differential cross section 
 for inelastic scattering to the $2^+$ state of $^{12}$C. The support points are at $E_{lab}=$~6 and
40~MeV. The separable representation describes both
         differential cross sections very well. In addition, it is 
         in good agreement with the coupled-channel calculations
         shown in Fig.~1 of Ref.~\cite{Olsson:1989npa}.  
The dashed lines indicate cross sections computed with the spherical
     Olsson~89~\cite{Olsson:1989npa} OMP.

\subsubsection{Off-shell $t$ matrices}
\begin{figure}
 \includegraphics[width=15cm]{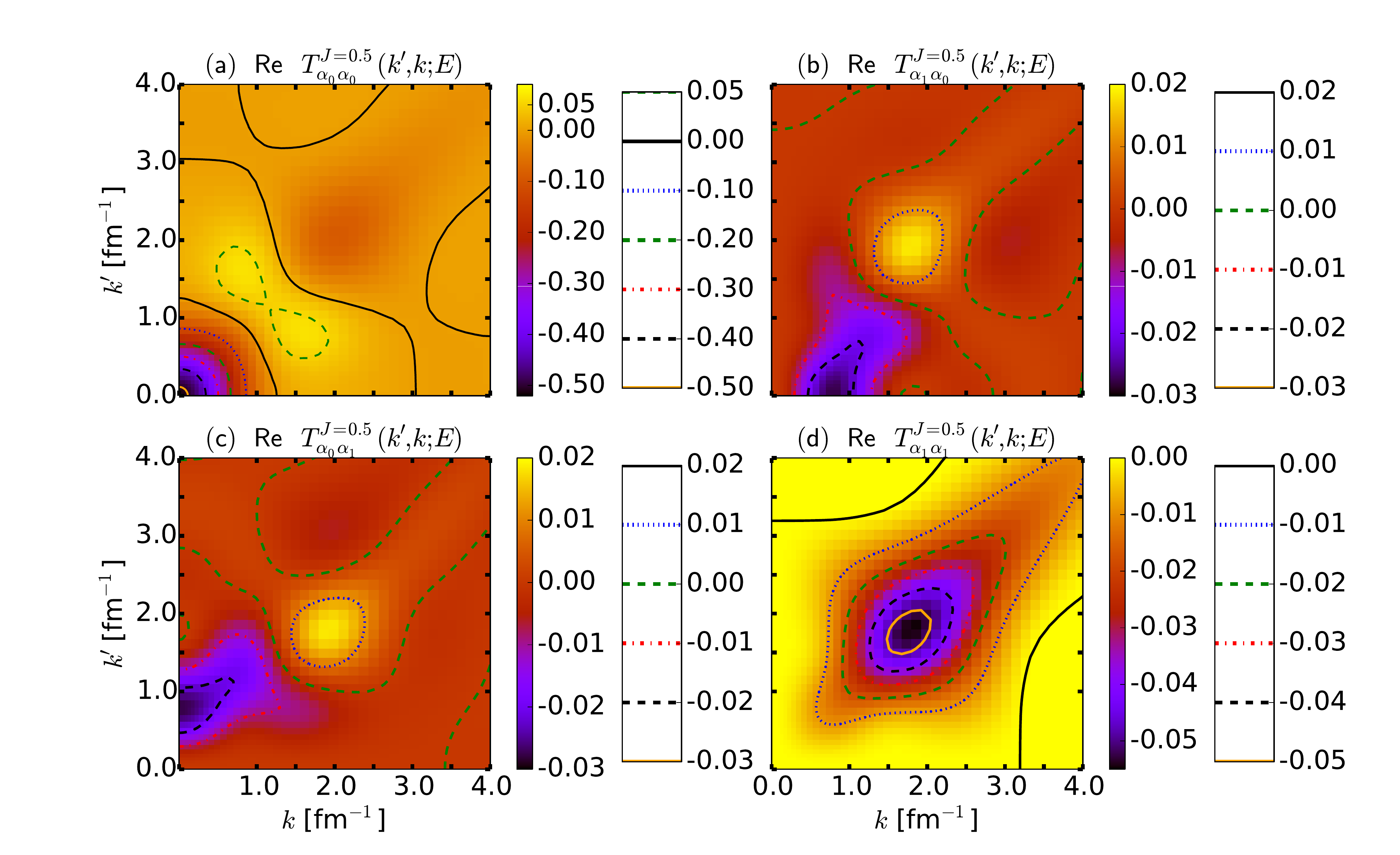}
 \caption{ The real part of the $t$ matrix elements for  $J^\pi=1/2^+$ 
  in the $n+^{12}$C system at 20.9~MeV incident neutron energy. 
  Panels (a), (b), (c), and (d) depict 
  $T_{\alpha_0\alpha_0}$, $T_{\alpha_1\alpha_0}$, $T_{\alpha_0\alpha_1}$,
  and $T_{\alpha_1\alpha_1}$. The quantum numbers
   for the channels shown here are $\alpha_0=\{I=0,l=0,j_p=1/2\}$, and $\alpha_1=\{I=2,l=2,j_p=1.5\}$.
The on-shell momentum is given by $k_0$~=~0.93~fm$^{-1}$. 
  \label{fig:fig4}
  }
  \end{figure}

 The reciprocity theorem requires that the off-shell $t$ matrix be invariant
under simultaneous interchange of channel indices and momenta.
 It is thus imperative that we investigate the properties of the
 off-shell $t$ matrix elements computed with the separable representation
 schemes. First, we calculate the  
 multichannel off-shell $t$ matrix with the  original Olsson~89
 DOMP and show the real parts of the $J^\pi=1/2^+$ $t$ matrix for the $n+^{12}$C system at 20.9~MeV
incident neutron energy in Fig.~\ref{fig:fig4}. 
  The matrix elements correspond to the quantum numbers $\alpha_0=\{I=0,l=0,j=0.5\}$
   and $\alpha_1=\{I=2,l=2,j=3/2\}$. We observe that
   the $t$ matrices exhibit high momentum components in all channels, which
    is characteristic of local potentials. In the coupled channels they 
   are invariant under
   the simultaneous interchange of channel indices
   and momenta, as required by the reciprocity theorem.
  
\begin{figure}
 \includegraphics[width=15cm]{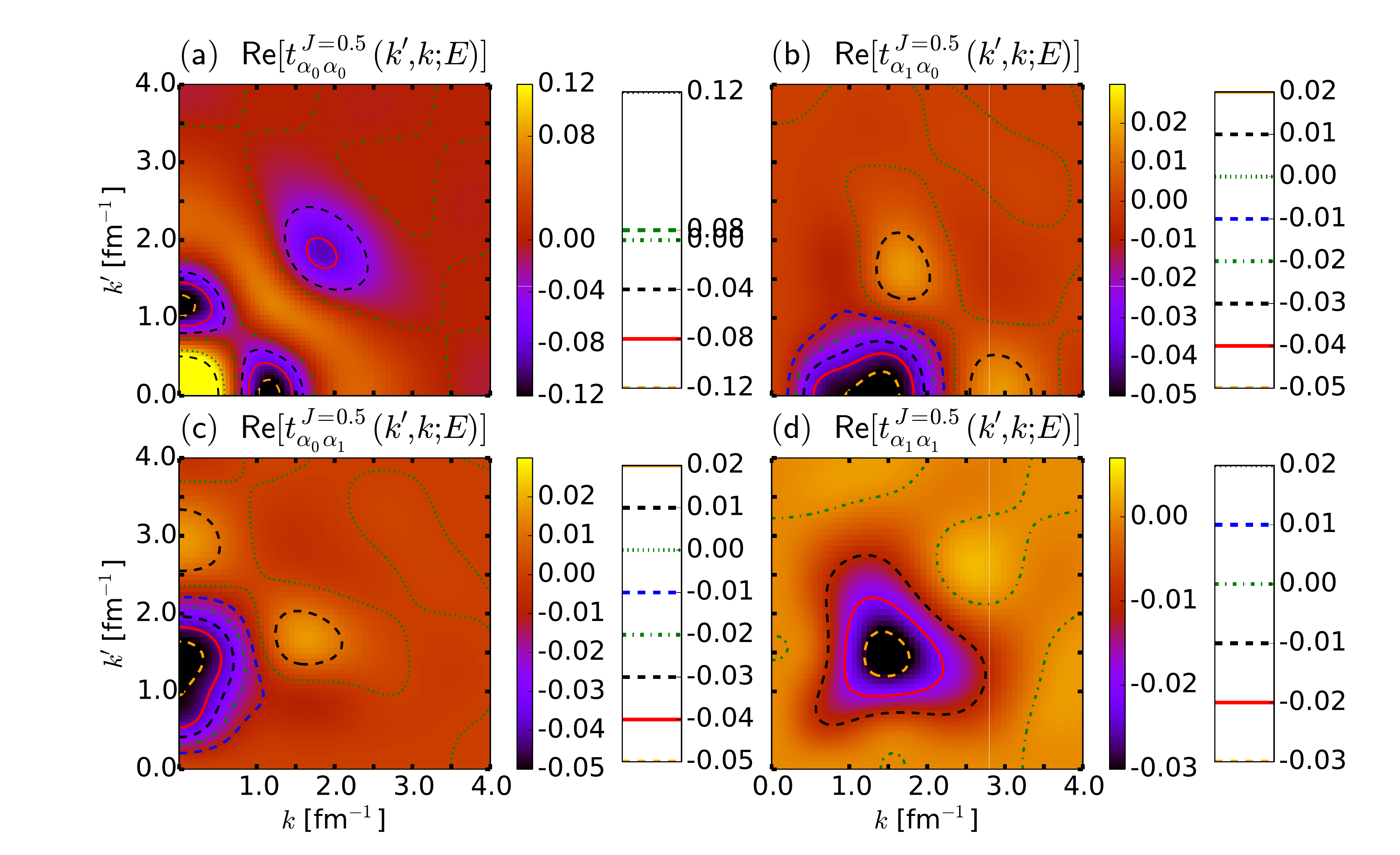}
 \caption{The  real part of 
 multichannel off-shell eEST separable $t$ matrix elements
 for the $n+^{12}$C system at 20.9~MeV
 incident neutron energy.  Panels (a), (b), (c), and (d) depict 
 $T_{\alpha_0\alpha_0}$, $T_{\alpha_1\alpha_0}$, $T_{\alpha_0\alpha_1}$,
  and $T_{\alpha_1\alpha_1}$.
  The quantum numbers for the states shown here are $J^\pi=1/2^+$
   $\alpha_0=\{I=0,l=0,j=0.5\}$, and $\alpha_1=\{I=2,l=2,j=1.5\}$.   
The on-shell momentum is given by $k_0$~=~0.93~fm$^{-1}$.
  \label{fig:fig5}
  }
  \end{figure}

\begin{figure}
 \includegraphics[width=15cm]{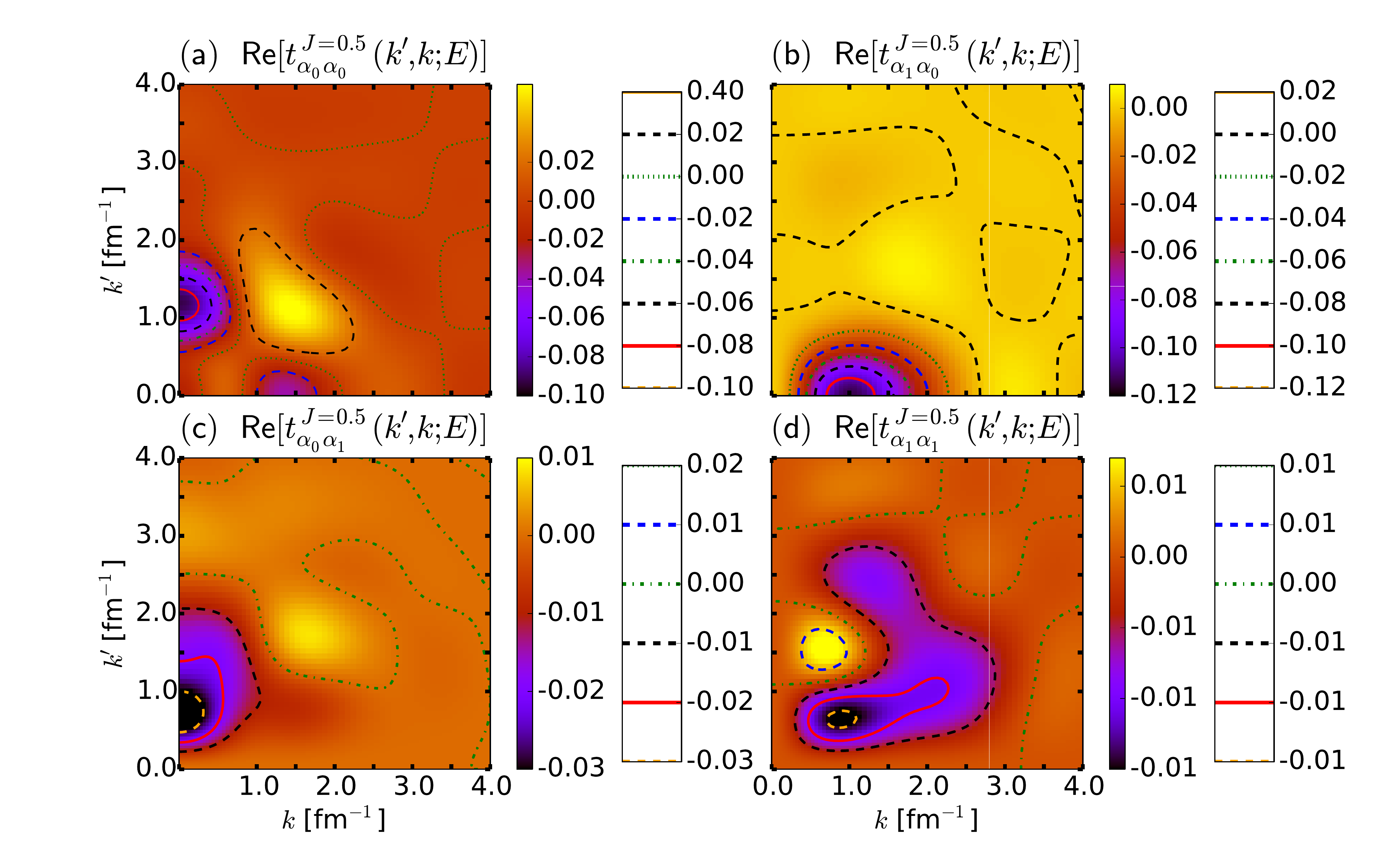}
 \caption{ Same as Fig.~\ref{fig:fig8} but for the EST separable 
   representation of the $t$ matrix.    
  \label{fig:fig6}
  }
\end{figure}

 Next, we explore the off shell properties of $t$ matrix obtained
 with the eEST separable representation scheme. In Fig.~\ref{fig:fig5}
 the real part of the  multichannel  eEST separable $t$ matrix is shown  
 for the $n+^{12}$C system at 20.9~MeV incident neutron energy for the same channels.
First we observe that the separable representation does not contain high-momentum components in either
channel. 
 We also see that $t$~matrix elements
 obtained with the eEST separable representation
 are invariant under a simultaneous
 interchange of channel indices and momenta, 
 as required by the reciprocity theorem. 
To illustrate that the energy-independent EST separable representation  is deficient in that 
respect, Fig.~\ref{fig:fig6} shows the off-shell $t$ matrix for the same channels. It is
quite
obvious that the channel-coupling $t$ matrices, panels (b) and (c), are not invariant under a
simultaneous interchange of channel indices and momenta. However, even the diagonal channels, panels
(a) and (d) are not symmetric under the exchange of $k$ and $k'$. This violation was already found
for single channel energy-independent EST calculations~\cite{Hlophe:2016}. In multi-channel
calculations this violation of symmetry seems to be enhanced.
 
\begin{figure}
 \includegraphics[width=17cm]{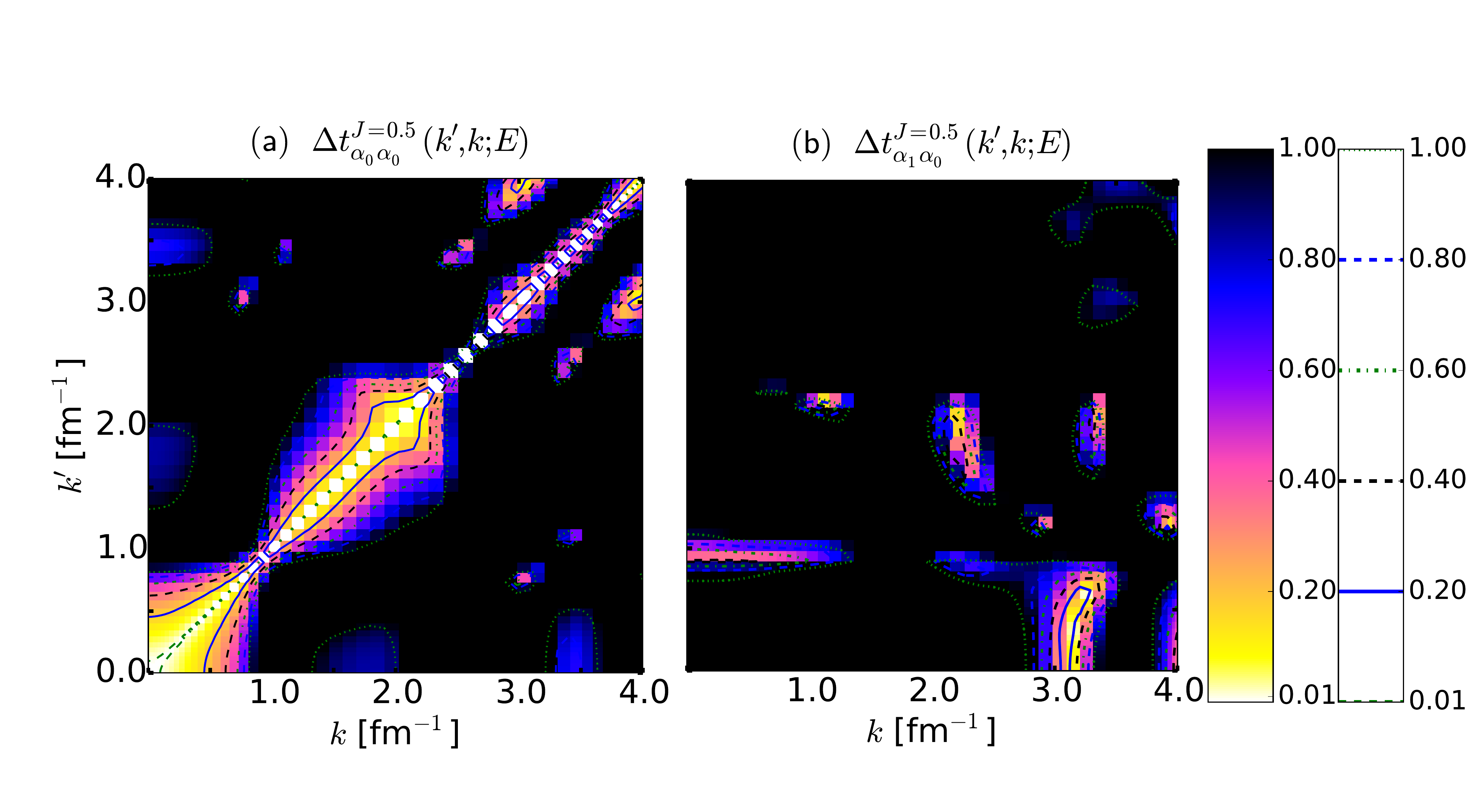}
 \caption{ The  $J^\pi=1/2^+$ asymmetry for the $n+^{12}$C system evaluated
    at 20.9~MeV as function of the  off-shell momenta $k'$ and $k$. 
   Panels (a)and (b) show the asymmetry for  the energy-independent EST 
    separable representations. 
  \label{fig:fig7}
  }
\end{figure}

To determine the extent to which reciprocity is violated (as in Ref.~\cite{Hlophe:2016}), we define an asymmetry relation as
 \begin{eqnarray}
   \Delta t_{\alpha\alpha_0}^J(k',k;E)=\left|\frac{t_{\alpha\alpha_0}^J(k',k;E)
-t_{\alpha_0\alpha}^J(k,k';E)} {[t_{\alpha\alpha_0}^J(k',k;E)+t_{\alpha_0\alpha}^J(k,k';E)]/2}\right|.
\label{eq:asym-mch}
\end{eqnarray}
 In Fig.~\ref{fig:fig7} this asymmetry is shown for the $n+^{12}$C system computed at
 20.9~MeV as function of  the off-shell momenta $k'$ and $k$ for the energy-independent EST
representation. Panel (b) clearly shows  
that even at the on-shell point ($k'=k=k_0=$~0.93~fm$^{-1}$) the asymmetry in the  $\alpha\ne\alpha_0$ channel is non-zero. 
This behavior can not be repaired by increasing the rank of the separable representation. 
The same calculation for the eEST representation will give exactly zero for all values of $k'$ and $k$
in all channels. The panels would be white and are therefore not shown. 
From this we conclude that for a separable representation of energy-dependent multi-channel  complex optical potential, 
the eEST scheme must be applied if reciprocity should be fulfilled.

\section{Separable representation of energy-dependent 
 Multichannel Proton-Nucleus Optical Potentials}
\label{mch-popt}

\subsection{Formal Considerations}
\label{popt-eest}

The interaction of protons with nuclei comprises of the strong
nuclear force and the Coulomb potential. The nuclear interaction
is given by the complex, energy-dependent optical potential  and usually 
has the same form as the neutron optical potential. 
The Coulomb interaction consists of
a long-ranged point-Coulomb potential $V^C$ and a short-ranged piece. The latter arises from the charge distribution of the nucleus and is commonly approximated by a uniformly charged sphere. 
The point-Coulomb potential is long-ranged and affects the asymptotic behavior of the scattering wavefunctions. Consequently, the proton-nucleus scattering
problem can not be treated with the same techniques employed for neutron scattering in Section~\ref{mch-nopt}. According to the
Goldberger-Gell-Mann relation~\cite{RodbergThaler}, the scattering amplitude for $p+A$ scattering separates into two parts. The first part is the Rutherford amplitude corresponding to the point Coulomb potential.
The second part is the nuclear amplitude corresponding to the short-range potential $U$, consisting
of the nuclear interaction and the short-range Coulomb potential.
While the Rutherford amplitude is known analytically, the nuclear amplitude must be evaluated numerically. The nuclear
amplitude is evaluated in a basis of Coulomb wavefunctions according to Eq.~(\ref{eq:pad4d}). When working in momentum space,
this is a very challenging task since the Coulomb wavefunctions are singular. We employ 
the techniques presented in Ref.~\cite{Elster:1993dv}, where the authors showed that, 
in the Coulomb basis, the
Coulomb-distorted nuclear $t$ matrix fulfills a LS-type equation of the same form as in
the basis of plane waves.
The momentum-space matrix elements of the potential in the Coulomb basis are obtained
via Fourier transform from 
coordinate space. For multichannel potentials, the nuclear $t$ matrix fulfills the coupled set of LS-type equations shown in Eq.~(\ref{eq:cmulti3g}).

 To derive a separable representation for the short-ranged
 proton-nucleus potential $U$, we modify the multichannel eEST scheme
 of Section~\ref{nopt-eest}. The multichannel scattering wavefunctions
 $\big|{\Psi}_{\rho,i}^{J\pi(+)}\big\rangle$  are
 replaced by the Coulomb-distorted multichannel scattering wavefunctions 
 $\big|{\Psi^c}_{\rho,i}^{J\pi(+)}\big\rangle$. This
 leads to the separable potential
 \begin{eqnarray}
  u(E)&=&\sum\limits_{\rho\sigma}\sum\limits_{ij} 
  \left(\sum\limits_{\gamma JM} U\big|\gamma JM\; {\Psi^c}_{\gamma\rho,i}^{J(+)}\big\rangle\right) \;
  {\lambda}_{ij} ^{c,\;\rho\sigma}(E)\;
 \left(\sum\limits_{\gamma JM}\big\langle {\Psi^c}_{\gamma\sigma,j}^{J(-)}\;\gamma JM\big|U\right),\cr\cr
 &=&\sum\limits_{JM}\sum\limits_{J'M'}\sum\limits_{\rho\sigma}\sum\limits_{ij} 
   {{T^c}}(E_i)\big|\rho JM\; {{\phi^c}}_i^\rho\big\rangle \;{\lambda}_{ij}
  ^{c,\;\rho\sigma}(E)\;
 \big\langle {{\phi^c}}_j^\sigma\;\sigma J'M'\big|{{T^c}}(E_j).
 \label{eq:cmch21a}
 \end{eqnarray}
Here ${{\phi^c}}_j^\sigma$ are the scattering Coulomb wavefunctions corresponding to the channel momentum
$k_j^\sigma$. The Coulomb-distorted nuclear transition matrix $T^c(E)$ fulfills
the LS equation
\begin{eqnarray}
T^c(E)=U+UG_C(E)T^c(E),
\label{eq:cmch21aa}
\end{eqnarray}
with $ G_C(E)=[E-H_0-V^C+i\epsilon]^{-1}$
 being the Coulomb Green's
function, $H_0$ the free Hamiltonian, and $V^C$ the point-Coulomb potential.  
 In analogy to Eqs.~(\ref{eq:mch21b0}) and (\ref{eq:mch21b1}), we
 define the energy-dependent matrix
 \begin{eqnarray}
  {\cal U}^{e,\alpha\beta}_{mn}(E)&\equiv&\sum\limits_{\gamma\nu}\big\langle {\Psi^c}_{\gamma\alpha,m}^{J(-)}\;\gamma JM\big|
  U(E)\big|\nu JM\;{\Psi^c}_{\nu\beta,n}^{J(+)}\rangle,\cr\cr
   &=&\sum\limits_{\gamma\nu}\big\langle {\Psi^c}_{\gamma\alpha,m}^{J(-)}\big|
   U_{\gamma\nu}^J(E)\big|{\Psi^c}_{\nu\beta,n}^{J(+)}\big\rangle,
   \label{eq:cmch21b0}
 \end{eqnarray}
and the energy-independent matrix
 \begin{eqnarray}
  {{\cal U}^c}^{\alpha\beta}_{mn}={\cal U}^{ce,\alpha\beta}_{mn}(E_m)&\equiv&\sum\limits_{\gamma\nu}\big\langle {\Psi^c}_{\gamma\alpha,m}^{J(-)}\;\gamma JM\big|
  U(E_m)\big|\nu JM\;{\Psi^c}_{\nu\beta,n}^{J(+)}\rangle,\cr\cr
   &=&\sum\limits_{\gamma\nu}\big\langle {\Psi^c}_{\gamma\alpha,m}^{J(-)}\big|
   U_{\gamma\nu}^J(E_m)\big|{\Psi^c}_{\nu\beta,n}^{J(+)}\big\rangle.
   \label{eq:cmch21b1}
 \end{eqnarray}
  The coupling matrix elements $\lambda_{ij}^{c,\;\rho\sigma}(E)$ then
  fulfill 
 \begin{eqnarray}
  {\cal U}^{ce,\alpha\beta}_{mn}(E)&\equiv&\sum\limits_{\gamma\nu}\big\langle {\Psi^c}_{\gamma\alpha,m}^{J(-)}\;\gamma JM\big|
  u(E)\big|\nu JM\;{\Psi^c}_{\nu\beta,n}^{J(+)}\big\rangle,\cr\cr
 &=&\sum\limits_{\rho\sigma}\sum\limits_{ij}\sum\limits_{\gamma\nu} 
   \big\langle {\Psi^c}_{\gamma\alpha,m}^{J(-)}\;\gamma JM\big|{T^c}(E_i)\big|\rho JM\; {{\phi^c}}_i^\rho\big\rangle \;\lambda_{ij}
  ^{c,\;\rho\sigma}(E)\;\cr\cr
 &\times& \big\langle {\phi^c}_j^\sigma\;\sigma JM\big|{T^c}(E_j)\big|\nu JM\;{\Psi^c}_{\nu\beta,n}^{J(+)}\big\rangle,\cr
 &=& \left[{{\cal U}^c}^t\cdot \lambda^c(E)\cdot{\cal U}^c\right]_{mn}^{\alpha\beta}.
  \label{eq:cmch21b2}
 \end{eqnarray}
Here ${{\cal U}^c}^t$ represents the transpose of ${\cal U}^c$.
The separable representation of the $t$ matrix is then given as
 \begin{eqnarray}
  t^c(E)&=&\sum\limits_{\rho\sigma}\sum\limits_{ij} 
  \left(\sum\limits_{\gamma JM} U\big|\gamma JM\; {\Psi^c}_{\gamma\rho,i}^{J(+)}\big\rangle\right) \;\tau_{ij}
  ^{c,\;\rho\sigma}(E)\;
 \left(\sum\limits_{\gamma JM'}\big\langle {\Psi^c}_{\gamma\sigma,j}^{J(-)}\;\gamma JM'\big|U\right),\cr\cr
 &=&\sum\limits_{JM}\sum\limits_{J'M'}
  \sum\limits_{\rho\sigma}\sum\limits_{ij} 
   {T^c}(E_i)\big|\rho JM\; {{\phi^c}}_i^\rho\big\rangle \;\tau_{ij}
  ^{c,\;\rho\sigma}(E)\;
 \big\langle {\phi^c}_j^\sigma\;\sigma J'M'\big|{T^c}(E_j).
 \label{eq:cmch21c}
 \end{eqnarray}
Substituting Eqs.~(\ref{eq:cmch21a})-(\ref{eq:cmch21c}) into the LS equation leads to
 \begin{eqnarray}
   R^c(E)\cdot\tau^c(E)\cdot{\cal U}^c=\mathcal{U}^{ce}(E),
   \label{eq:cmch24a}
 \end{eqnarray}   
 where
 \begin{eqnarray}
  R^{c,\;\rho\sigma}_{ij}(E)&=&\Big\langle {\phi^c}_i^{\rho}\Big|~T_{{\rho\sigma}}^{cJ}(E_i)~+~\sum\limits_{{\beta}}
  T_{{\rho\beta}}^{cJ}(E_i)G_{{C\beta}}(E_j)T_{ {\beta\sigma}}^{cJ}(E_j)\Big|\phi_j^{c\sigma}
  \Big\rangle\cr\cr
 &-&~\sum\limits_{ {\beta\beta}'}\sum\limits_{n}{\cal M}_{in}^{c,\;\rho\beta}(E)\langle  {\phi^c}_n^{\beta}\Big|
  T_{ {\beta\beta'}}^{cJ}(E_n)G_{ {C\beta'}}(E)T_{ {\beta'\sigma}}^{cJ}(E_j)
 \Big|\phi_j^{c\sigma}\Big\rangle,
   \label{eq:cmch24b}
 \end{eqnarray} 
 and
 \begin{equation}
 {\cal M}^{c,\;\rho\sigma}_{ij}(E)=\left[{\cal U}^{ce}(E)\cdot({\cal U}^c)^{-1}\right]^{\rho\sigma}_{ij}.
 \label{eq:mch25c-1}
 \end{equation}
Here the indices $\beta, \beta'$ represent angular momentum channels. 
  It is noteworthy that the expressions for the Coulomb distorted separable multichannel
  nuclear $t$ matrix elements have the same form as the ones obtained for
  neutron-nucleus systems in Section~\ref{nopt-eest}. The only difference is
 that  all quantities are evaluated in the Coulomb-basis. Moreover, the 
 behavior of the off-shell nuclear $t$ matrices under a transposition does
 not depend on the chosen basis, and thus the  
multichannel eEST separable representation for proton-nucleus systems fulfills the reciprocity in the same
fashion as the one for neutron-nucleus systems.

\subsection{ Elastic and Inelastic Scattering of Protons from $^{12}$C} 
\label{eest-pc12}

 To illustrate the implementation of the multichannel separable expansion presented in Section~\ref{popt-eest},
 we consider the scattering of protons from $^{12}$C including both the the $0^+$ ground state and $2^+$ excited state. A rigid rotor model is adopted to describe the structure of the $^{12}$C nucleus as in Section~\ref{eest-nc12}.
The proton-$^{12}$C interaction is given by a deformed OMP plus the point-Coulomb
 potential. The proton OMP includes a short-range contribution arising from the nuclear charge distribution. Deformations are introduced using multipole expansions as outlined in Appendix~\ref{domps}. In this work,
 we employ the Meigooni~85~\cite{Meigooni:1985iwq} DOMP which is presented in Appendix~\ref{meigooni84}. A uniformly charged sphere is assumed for the
 nuclear charge distribution so that the short-ranged Coulomb potential
 has the form
 \begin{eqnarray}
V_{coul}(r)= Z\alpha\Bigg[\frac{1}{2R_c}(3-r^2/R_c^2)-\frac{1}{r}\Bigg].      
 \label{eq:opt1a0}
\end{eqnarray}
The values of the Coulomb radius $R_c$ are adopted from Ref.~\cite{Weppner:2009qy}. Here
$Z$ is the atomic number and $\alpha$ the electromagnetic coupling constant.

\begin{figure}
 \includegraphics[width=15cm]{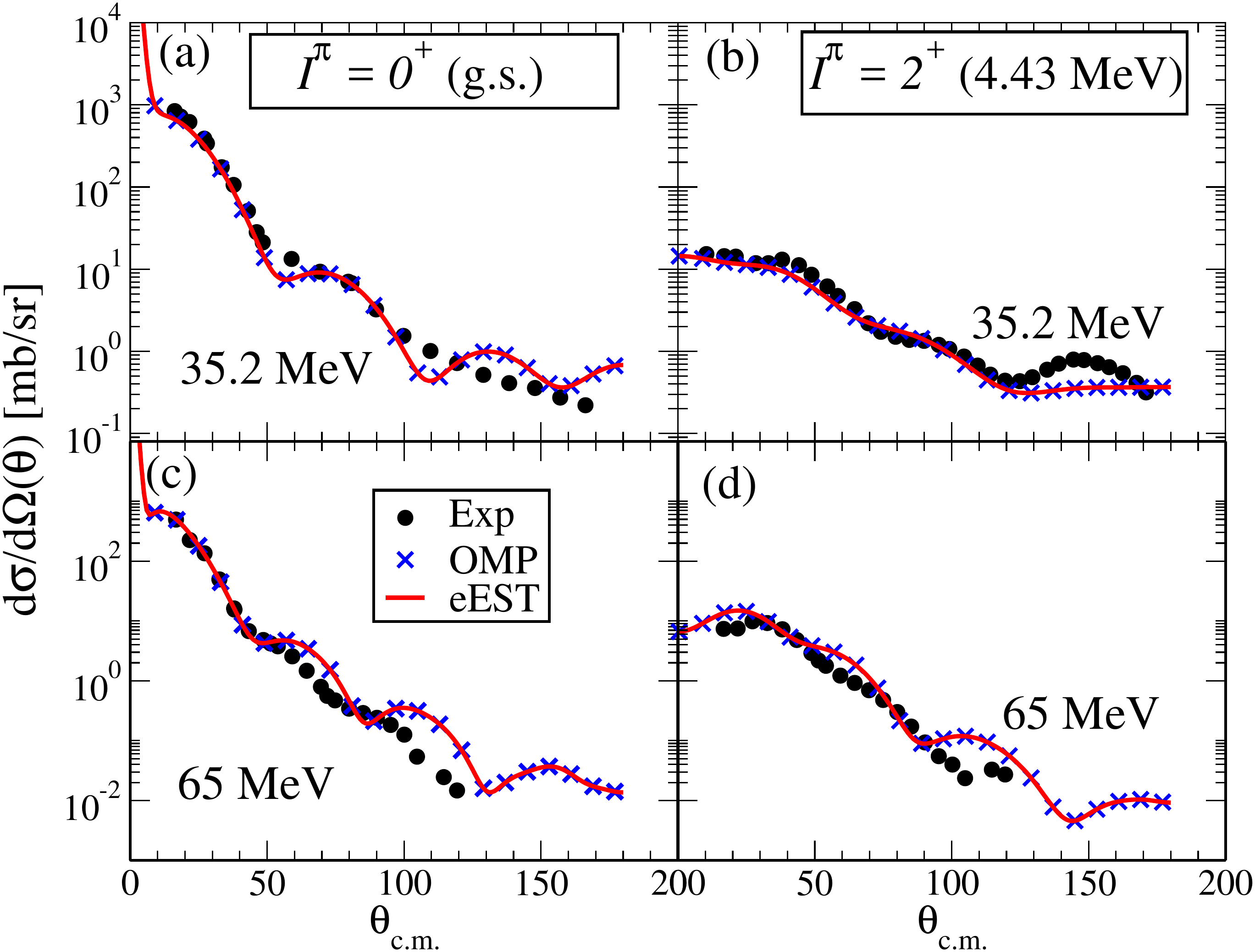}
 \caption{ The differential cross sections for proton scattering off $^{12}$C. Panels (a) and (c) depict the
 differential cross sections for elastic scattering 
 at proton incident energies of 35.2~MeV (a) and 65~MeV (c). 
 The corresponding cross sections
 for inelastic scattering to the $2^+$ state at 4.43~MeV are depicted in panels (b) and (d). 
The solid lines represent the calculations with the eEST separable representation, while calculations with
the original optical potential by Meigooni~85~\cite{Meigooni:1985iwq} are given by the crosses.
The support points for the separable representation at located at
 $E_{lab}=$~25, 45, and 65~MeV.
The data represented by filled circles at 35  and 65~MeV are
 taken from Refs.~\cite{Fabrici1980} and ~\cite{Kato1985} respectively.       
  \label{fig:fig8}
  }
\end{figure}

First, the differential cross sections are evaluated using the Meigooni~85 DOMP
and compared to the ones obtained using its separable representation. 
In Fig.~\ref{fig:fig8} the differential cross section
for elastic proton scattering from $^{12}$C  as function of the center
of mass (c.m.) angle $\theta_{c.m.}$ is shown for proton incident energies at 35.2~MeV (panel (a)) and 
65~MeV (panel (c)).  The solid lines show  the eEST separable representation while the
crosses represent calculations based on the original Meigooni~85 DOMP. The support points
for the separable representation are chosen to be at $E_{lab}=$~25, 45, and 65~MeV.
The differential cross sections for inelastic scattering  to the $2^+$ state are shown in 
panels (b) and (d) for the same energies. 
We observe that the cross sections computed with the  eEST separable representation of rank-3 agree 
very well with the ones computed directly from the Meigooni~85 DOMP.   

\begin{figure}
 \includegraphics[width=15cm]{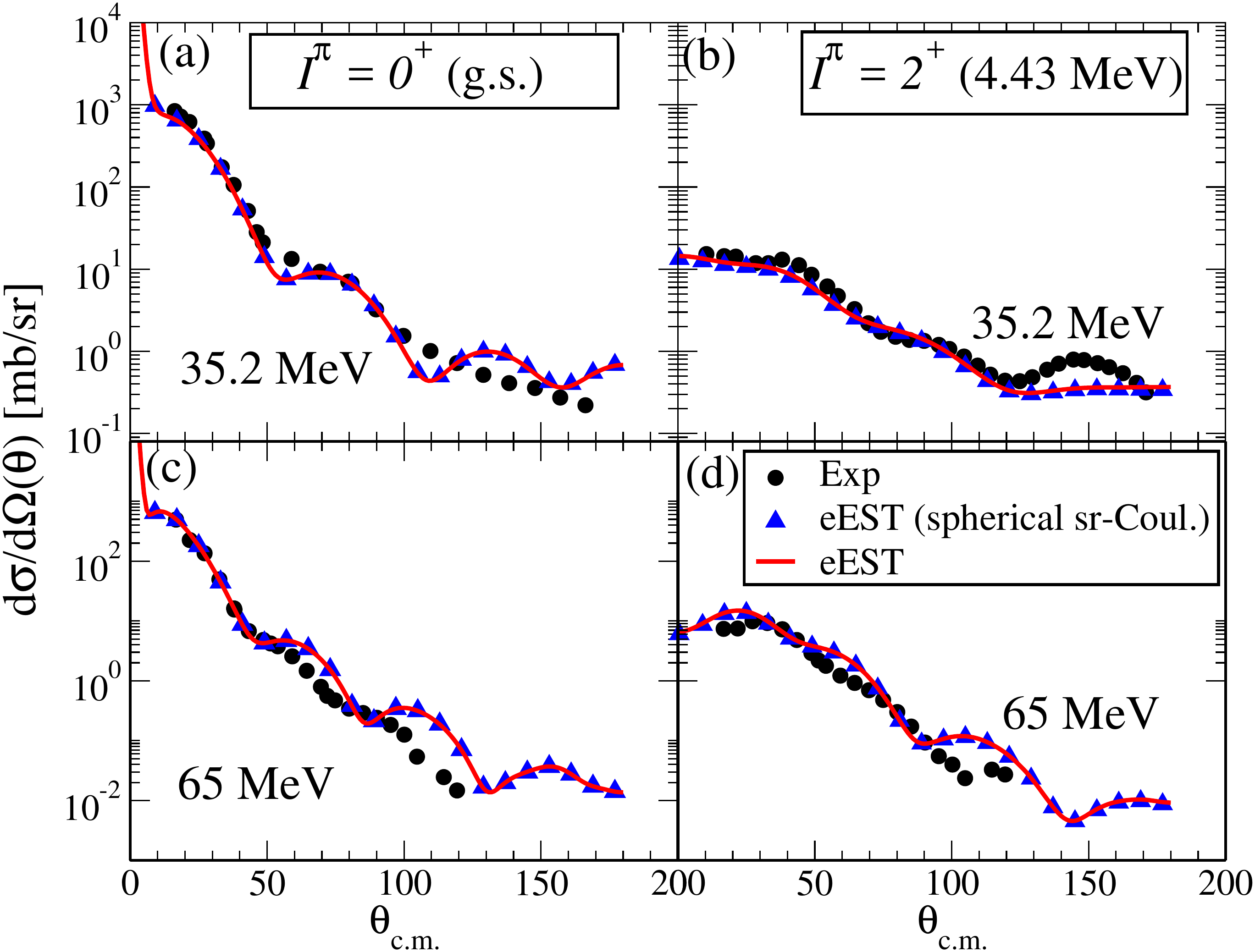}
 \caption{The differential cross sections for proton scattering off $^{12}$C. Panels (a) and (c) depict the
 differential cross sections for elastic scattering 
 at proton incident energies of 35.2~MeV (a) and 65~MeV (c).
 The corresponding cross sections
 for inelastic scattering to the $2^+$ state at 4.43~MeV are depicted in panels (b) and (d). 
The solid lines represent the calculations with the eEST separable representation. The support points for
the separable representation at located at
$E_{lab}=$~25, 45, and 65~MeV. The filled triangles represent a calculation in which only the short range
nuclear potential is deformed, while the short-ranged Coulomb potential spherical.
 The data at 35  and 65~MeV
  are aken from Refs.~\cite{Fabrici1980} and ~\cite{Kato1985} respectively.    
  \label{fig:fig9}
  }
   \end{figure}

When incorporating the rotational excitation of  $^{12}$C in the calculations presented in
Fig.~\ref{fig:fig8}, we deformed the nuclear part of the optical potential as well as the short-ranged
Coulomb potential, since the charge distribution should undergo the same deformation as the nuclear
short-ranged potential. In order to investigate if deforming the short-ranged Coulomb
potential affects the cross sections, we carry out the same calculations as before, but keeping the
short-range Coulomb potential spherical. The result of this calculation is shown in
Fig.~\ref{fig:fig9} by the filled upward triangles. It is interesting to note that the differential cross
sections for inelastic scattering do not show any effect of this simplification. 
This may be due to the still relatively small charge (Z=6) of $^{12}$C where a deviation from a
spherical charge has a small effect. In the framework of (d,p) reactions on $^{24}$Mg it was shown
that for Z=12 the deformation of the charge only leads to a very small effect in the transfer cross
section~\cite{Deltuva:2016muy}, thus our finding is consistent.
However,
since the strength of the short-ranged Coulomb force depends
on nuclear charge, its deformation would have a larger effect
on the cross sections for heavier nuclei.

\begin{figure}
 \includegraphics[width=15cm]{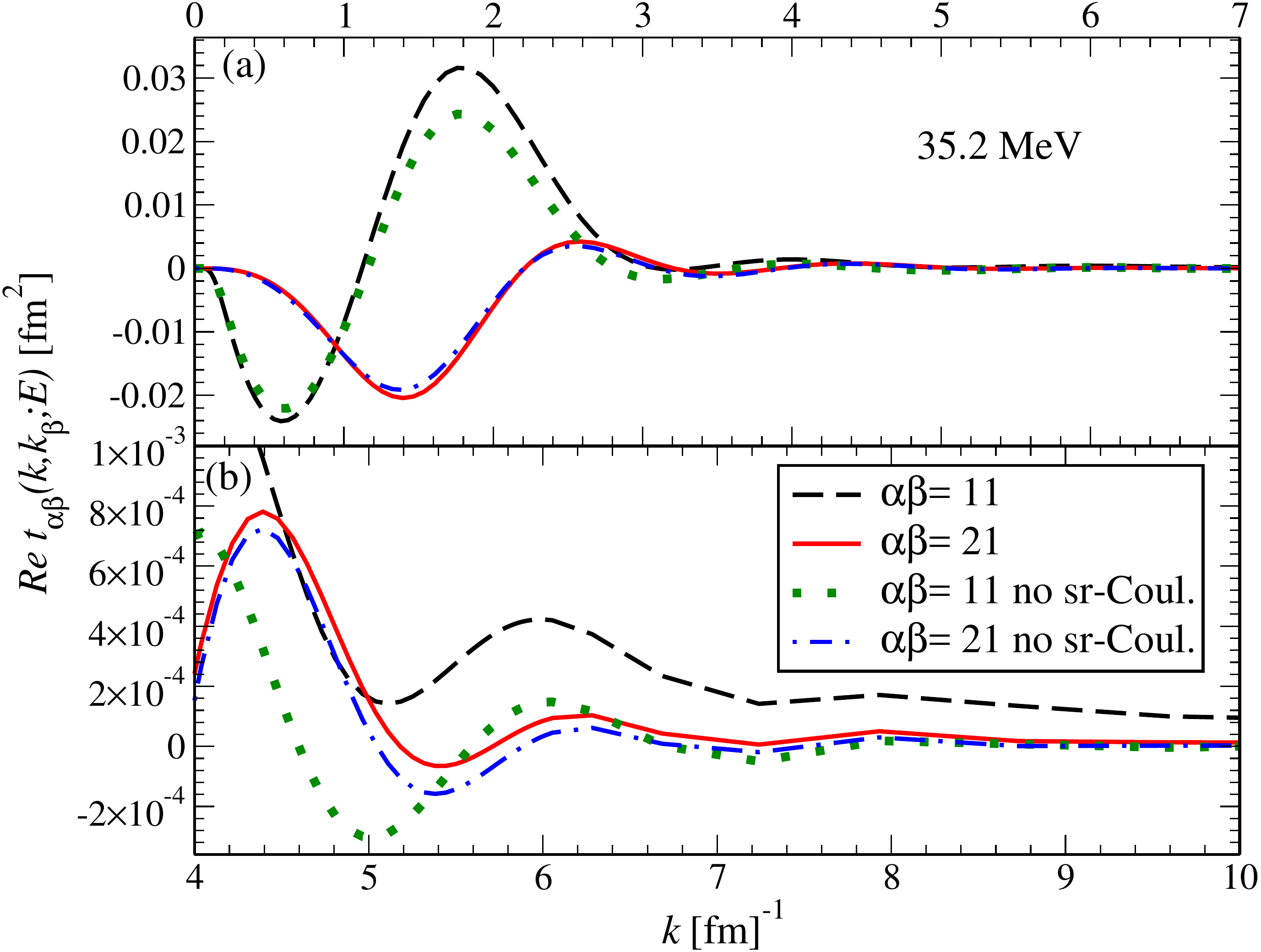}
 \caption{ The real part of  half-shell multichannel $t$ matrix elements
   $t_{\alpha\beta}^{J^\pi}(k,k_\beta;E)$ for $J^\pi=1/2^+$ for proton scattering from $^{12}$C
  at incident proton energy 35.2~MeV. Panel (a) shows the half-shell $t$ matrix in the interval
$0$~fm$^{-1}\le k\le 7$~fm$^{-1}$, while  panel (b) depicts the same half-shell $t$ matrix between
  $k= 4$ and 10~fm$^{-1}$. 
The  channels `$1$' and `$2$' are represented as
$1 \equiv \{I=0,\;l=0\;,j=1/2\}$ and  $2 \equiv \{I=0,\;l=2\;,j=3/2\}$. 
The dashed line represents the $t$ matrix elements $t_{11}^{1/2^+}(k,k_1;E)$ calculated from the
eEST separable representation of the Meigooni~85 OMP~\cite{Meigooni:1985iwq} for $k_1=1.2$~fm$^{-1}$
as function of $k$. The solid line gives the the $t$ matrix elements $t_{21}^{1/2^+}(k,k_1;E)$ obtained
in the same fashion,  but multiplied by a factor 3 to enhance its features.
For the calculations of the same channels, 
represented with the dotted and dash-dotted lines the short-ranged Coulomb
potential is omitted.
\label{fig:fig10}
  }
\end{figure}

Differential cross sections are summed up over all partial waves. The fact that the eEST scheme represents the cross sections directly obtained from the Meigooni~85 DOMP very well 
implies that overall the partial waves must be well represented. 
To study further details we now concentrate on the $J^\pi = 1/2^+$ state and study the half $t$
matrices $t^{J^\pi}_{\alpha,1} (k,k_1;E)$  calculated at the incident proton energy 35.2~MeV 
in more detail. Here $\alpha$ can take the values
$\alpha=1=\{I=0,\;l=0\;,j=1/2\}$ and $\alpha=2=\{I=0,\;l=2\;,j=3/2\}$. The real parts of the $t$
matrices in those channels are depicted in Fig.~\ref{fig:fig10}.
 Panel (a) shows the half-shell $t$ matrix elements for the channels `11' and `21' 
 in the interval $0$~fm$^{-1}\le k\le 7$~fm$^{-1}$ for the full calculation and a calculation
in which the short-range Coulomb potential is omitted. The matrix elements in the `21' channel are
multiplied with a factor 3 to be roughly of the same size as the matrix elements in the `11' channel.
  Since the fall-off behavior of the matrix
elements for large momenta may be important for reaction calculations, we depict in panel (b) the
matrix elements in the momentum interval $4$~fm$^{-1}\le k\le 10$~fm$^{-1}$. Here we see that the
short-range Coulomb potential mainly influences the diagonal `11' channel. At 6~fm$^{-1}$ the 
$t$ matrix calculated with the nuclear potential only is essentially zero, while the short-ranged
Coulomb potential still gives a small contribution. In the `21' channel both $t$ matrices 
fall off to
zero, indicating that the deformation of the short-ranged Coulomb potential has little effect for a
nucleus as light as $^{12}$C. This may however be different when considering heavier nuclei and will
have to be investigated further.

\begin{figure}
\includegraphics[width=15cm]{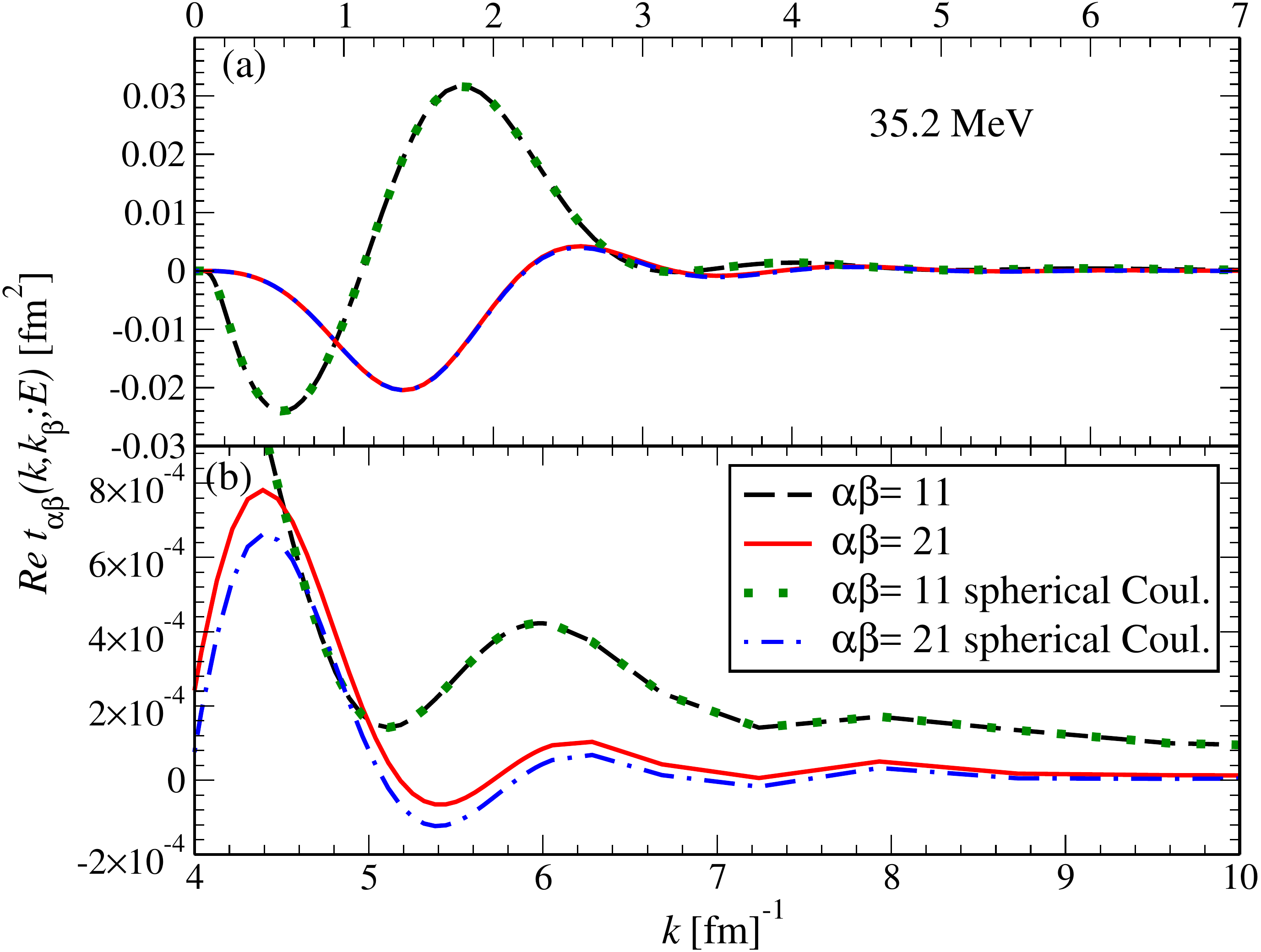}
\caption{ 
Same as Fig.~\ref{fig:fig10}. However for the calculations 
  represented with the dotted and dash-dotted lines the short-ranged Coulomb 
  potential is included but taken to be spherical. The t-matrix elements of the coupling channels are
multiplied by a factor 3 to enhance their features. 
\label{fig:fig11}
  }
   \end{figure}

 Finally, we separately investigate the effects of the deformation of the short-ranged Coulomb
potential on the half-shell multichannel $t$ matrix elements. To do so, 
 the coupled-channel calculation is carried out with a spherical
 short-ranged Coulomb potential and compared with the full calculation. 
This comparison is shown in Fig.~\ref{fig:fig11}. We plot the $t$ matrices in the same 
momentum ranges as in Fig.~\ref{fig:fig10}. In the curves in panel (a) we can not discern between
the two calculations. Only when considering a much smaller scale in panel (b) the calculations 
in the `21' channel are slightly different. However, the differences are so small that 
using a spherical short-range Coulomb potential can be considered a good approximation for $^{12}$C,
as already seen in the cross section at high momentum transfer in Fig.~\ref{fig:fig9}.

 \section{Summary and Conclusion}
 \label{summary}

In this work we introduce separable representations of complex, energy-dependent
multichannel optical potentials for neutron as well as proton scattering from
nuclei. To fulfill reciprocity exactly, the separable representation must 
be energy-dependent. This is achieved by having energy-dependent coupling constants.

 The first part of the manuscript concentrates on the separable 
 expansion of neutron-nucleus Deformed Optical Model Potentials (DOMPs). 
This is achieved by generalizing the energy-dependent 
 EST (eEST) scheme of Ref.~\cite{Hlophe:2016} to multichannel potentials.
 To illustrate the implementation of the multichannel eEST scheme, we considered
 neutron scattering from $^{12}$C.  In order to 
describe the structure of the $^{12}$C nucleus, the rigid rotor model is
assumed, and the Olsson~89~\cite{Olsson:1989npa} DOMP was adopted to describe 
the effective neutron-$^{12}$C interaction.        
 The multichannel eEST scheme  was then used to construct
 a separable representation for the Olsson~89 DOMP. In this case 
a rank-2 separable expansion was sufficient 
to describe elastic and inelastic scattering cross sections 
between 0 and 50~MeV incident neutron energies.

To demonstrate the necessity of using an energy-dependent multichannel separable 
representation, we also constructed 
a separable expansion based on the energy-independent
  multichannel EST scheme, which needed to be of rank-3 to match the 
  representation of $S$-matrix elements between 0~and~50~MeV with the same
quality as the eEST scheme.
An examination of the  off-shell $t$ matrix elements
  showed that only the multichannel eEST separable representation
  fulfills reciprocity, i.e. the $t$ matrix is invariant upon a simultaneous interchange
  of momenta and channel indices. In contrast, the energy-independent multichannel
expansion yields an asymmetric off-shell $t$ matrix, as was already observed in single
channel expansions in Ref.~\cite{Hlophe:2016}.

The cost of implementing the eEST scheme for multichannel
potentials increases with the number of channels, since the 
matrix elements ${\cal U}^{e,\alpha\beta}_{ij}(E)$ must be calculated at
each desired energy $E$. However, it turns out that it is sufficient to 
compute this matrix element at fixed energies and interpolate on the energy to obtain
its value at arbitrary energies. As already observed in Ref.~\cite{Hlophe:2016}
  for single-channel potentials, the results obtained with the 
  interpolated eEST scheme agree very well with the ones obtained without
  the interpolation.

The second part of the manuscript focuses on the separable representation
of proton-nucleus DOMPs. 
The proton-nucleus potential consists of a nuclear piece as well as the Coulomb interaction. 
The Coulomb interaction further separates into a short-ranged
part, usually represented as a charged sphere,  and a the long-ranged point-Coulomb force. The 
point-Coulomb potential is incorporated by working in the
Coulomb basis in accordance with the Gell-Mann-Goldberger~\cite{RodbergThaler}
relation. In order to employ the eEST scheme in the same fashion as for neutron-nucleus
scattering, we need to solve a Lippmann-Schwinger type equation to obtain 
half-shell $t$ matrices in the Coulomb basis.
While the Coulomb propagator is quite simple in this basis,
the evaluation of  potential matrix elements is more involved. 
We followed Ref.~\cite{Elster:1993dv} to evaluate the potential matrix elements.

To demonstrate an implementation of an eEST representation of a proton-nucleus optical
potential, we considered proton scattering off $^{12}$C and used the
Meigooni~85~\cite{Meigooni:1985iwq} DOMP as a starting point.
  Differential
 cross sections for elastic and inelastic scattering were  computed using the
eEST scheme, showing that
a rank-3 separable expansion is sufficient to represent the
Meigooni~85~\cite{Meigooni:1985iwq} DOMP.
Note that the only difference between the eEST schemes for neutron and proton
optical potentials is the basis employed for the separable expansion.
This implies that the discussions on reciprocity given in
Section~\ref{eest-nc12} for neutron-nucleus potentials apply
to proton-nucleus systems as well.
   
In the Meigooni~85 DOMP the short-ranged Coulomb potential is deformed in the 
same fashion as the short-range nuclear potential.
To study the
effects of deforming the short-ranged Coulomb
potential, the coupled-channel calculations were repeated
with the Coulomb deformation parameter set to zero.
It was observed that the differential cross sections are
not significantly altered by using a spherical short-ranged
Coulomb potential. The effect on the half-shell $t$ matrix elements for
elastic scattering is also negligible. There is a
minimal change to the half-shell $t$ matrix
elements  for inelastic scattering. However,
those changes are  so small that when considering a nucleus as light as $^{12}$C
a deformation of the charge distribution may be safely neglected. This insight is
consistent with the finding in Ref.~\cite{Deltuva:2016muy}.  Most likely, for heavier nuclei
this will not be the case and a deformation of the short-ranged Coulomb potential will
be mandatory.


\vfill


\begin{acknowledgments}
This work was performed in part under the
auspices of the U.~S.  Department of Energy under contract
No. DE-FG02-93ER40756 with Ohio University.
The authors thank F.M. Nunes 
for thoughtful comments and careful reading of the manuscript.
\end{acknowledgments}


\bibliography{lhlophe_tad-1.bib,lhlophe_tad-2.bib,lhlophe_tad-3.bib}

\clearpage

\clearpage

\appendix
\section{Nucleon Scattering from  a Deformed Nucleus}
\label{appx-A}

\subsection{Neutrons}
\label{appx-nascat}

Let us consider a neutron scattering from a nucleus possessing a rotational energy spectrum
 characterized by the spin-parity $I^\pi$. The spectrum corresponds to collective
 rotational states with wavefunctions
  $|\Phi_{IM_{I}}\rangle$. These are given by~\cite{Bohr:1998} 
  \begin{eqnarray}
  \Phi_{IM_IK}(\xi)=\left(\frac{2I+1}{8\pi^2}\right)^{1/2}{\cal D}_{MK}^I(\xi),
  \label{eq:rot3a}
  \end{eqnarray}
   where ${\cal D}_{MK}^I(\xi)$ are the Wigner rotation matrices.
   Here $\xi$ are the angles specifying the orientation of the nucleus. 
   The interaction between the neutron and the nucleus
 leads to couplings between states of different spin-parity. Here
 we treat couplings to selected rotational states
 explicitly, while the imaginary part of the optical potential accounts for couplings
 to all other channels besides those that are included. To compute scattering
 observables for such systems, the coupled-channels formalism~\cite{Bohr-fys:1953} must be adopted.  

 \subsubsection{States of conserved angular momentum}
 \label{jcouple}
 
 To construct states of conserved angular momentum we adopt the so-called
 $jj$-coupling scheme. The neutron spin $s=1/2$ is coupled
 to the relative orbital angular momentum $l$ to yield $j_p=|l\pm 1/2|$. 
 The corresponding projections along the $z$-axis $m_s$ and $m_l$ fulfill
 the condition $m_{j_p}=m_s+m_l$.
 The angular momentum $j_p$ is in turn coupled to the nuclear spin $I$ to yield
 the total angular momentum $|I-j_p|\le J\le I+j_p$. The angular momentum
 $J$ and its projection along the $z$-axis $M=M_I+m_{j_p}$ are conserved. 
 To obtain the states of conserved angular momentum $|(Ilsj_p)JM\rangle$, we first define
 \begin{equation}
  |\mathcal{Y}_{ls}^{j_pm_{j_p}}\rangle= \sum\limits_{m_lm_s} C(lsj_p,m_lm_sm_{j_p})|Y_{lm_l}\rangle|sm_s\rangle,
  \label{eq:nad0a}
 \end{equation}
 where the functions $Y_{lm_l}$  are the spherical harmonics and $\chi_{sm_s}$ is the corresponding spinor.
 The quantity $C(lsj_p,m_lm_sm_{j_p})$ is the 
 Clebsch-Gordon (C.G.) coefficient. The inverse relation is given by
  \begin{equation}
 |Y_{lm_l}\rangle|{sm_s}\rangle= \sum\limits_{j_pm_{j_p}} C(lsj_p,m_lm_sm_{j_p})|\mathcal{Y}_{ls}^{j_pm_{j_p}}\rangle.
  \label{eq:nad0b}
 \end{equation}

The sates $|I(lsj_p)JM\rangle$ are  constructed by coupling Eq.~(\ref{eq:nad0a}) to the rotational state $|\Phi_{IM_I}\rangle$ so that
  \begin{eqnarray}
  |(Ilsj_p)JM\rangle = \sum\limits_{M_Im_{j_p}} C(Ij_pJ,M_Im_{j_p}M)|{\cal Y}_{ls}^{j_pm_{j_p}}\rangle|\Phi_{IM_{I}}\rangle,
 \label{eq:nad1a}
 \end{eqnarray}
 with inverse relation
  \begin{eqnarray}
  |{\cal Y}_{ls}^{j_pm_{j_p}}\rangle|\Phi_{IM_{I}}\rangle = \sum\limits_{JM} C(Ij_pJ,M_Im_{j_p}M)|(Ilsj_p)JM\rangle.
 \label{eq:nad1b}
 \end{eqnarray}
 Substituting Eq.~(\ref{eq:nad0a}) and~(\ref{eq:nad0b}) 
 into Eqs.(\ref{eq:nad1b}) and (\ref{eq:nad1a}) leads to
   \begin{eqnarray}
  |(Ilsj_p)JM\rangle = \sum\limits_{M_Im_{j_p}}\sum\limits_{m_lm_s} C(Ij_pJ,M_Im_{j_p}M) 
  C(lsj_p,m_lm_sm_{j_p})|Y_{lm_l}\rangle|sm_s\rangle|\Phi_{IM_{I}}\rangle,
 \label{eq:nad1c}
 \end{eqnarray}
 with the inverse relation given by
  \begin{eqnarray}
   |Y_{lm_l}\rangle|sm_s\rangle|\Phi_{IM_{I}}\rangle= \sum\limits_{JM}\sum\limits_{j_pm_{j_p}} 
   C(Ij_pJ,m_Im_{j_p}M) C(lsj_p,m_lm_sm_{j_p})|(Ilsj_p)JM\rangle.
 \label{eq:nad1d}
 \end{eqnarray}
 To simplify the notation, we define a channel index $\alpha\equiv(Ilsj_p)$ so that
 the states of conserved angular momentum can be written as $|(Ilsj_p)JM\rangle=|\alpha JM\rangle$.
 These states form a complete basis
 \begin{equation}
\mathbf{1}= \sum\limits_{\alpha JM} \alpha {JM}\rangle
 \langle\alpha JM|.
 \label{eq:nad1e}
\end{equation}

 \subsubsection{Scattering amplitudes and cross sections}
 \label{namch:obsv}
 
The on-shell $t$ matrix is related
  to the $S$-matrix by
  \begin{eqnarray}
   &&S_{\alpha\alpha_0}^J(E)= \delta_{\alpha\alpha_0}+2i\kappa_{\alpha\alpha_0}^J(E),\cr\cr
   &&\text{with~~~~} \kappa_{\alpha\alpha_0}^J(E)=-\pi\sqrt{\rho_\alpha\rho_{\alpha_0}}
   \;T_{\alpha\alpha_0}^J(k_0^\alpha,k_0^{\alpha_0};E) .
   \label{eq:nad3d}
  \end{eqnarray}
  Here $\rho_\alpha=\mu_{\alpha}k_0^\alpha$ is the density of states in channel $\alpha$.   
 To obtain the scattering amplitude for the process $ |{\bf k_0} sm_s{IM_I}\rangle
 \longrightarrow|{\bf k_\alpha} sm_s'{I'M_I'}\rangle$, we first evaluate the 
 $t$ matrix
\begin{eqnarray}
 T_{I'M_I'sm_s':IM_Ism_s}({\bf k_{\alpha}},{\bf k_{0}};E)&=&
 \langle {I'M_I'}sm_s'{\bf k_\alpha}|T(E)|{\bf k_0} sm_s{IM_I}\rangle\cr\cr
 &=&\sum\limits_{JM} \sum\limits_{ll'j_pj_p'}
C(Ij_pJ,M_I M-M_I M)\;C(lsj_p,M-M_I-m_s m_s M-M_I)\cr
 &\times&\;C(l'sj_p,M-M_I'-m_s' m_s' M-M_I')C(I'j_p'J,M_I' M-M_I' M)\cr
 &\times& T_{\alpha\alpha_0}^{J}(k_\alpha,k_0;E)\;Y_{l'M-M_I'-m_s'}({\bf\hat k_\alpha}) Y_{lM-M_I-m_s}^*({\bf\hat k_0}).
 \label{eq:nad4b}
 \end{eqnarray}
 The scattering amplitude for can be inferred from the relation~\cite{Joachain}
 \begin{eqnarray}
  f_{I'M_I'sm_s':IM_Ism_s}(E,\theta)&=&\langle{I'M_I'}sm_s'{\bf k_\alpha}|f(E)|{\bf k_0} sm_s{IM_I}\rangle\cr
  &=&\frac{4\pi^2}{k_{0}}\sqrt{\rho_{\alpha_0}\rho_\alpha}
  \langle{I'M_I'}sm_s'{\bf k_\alpha}|T(E)|{\bf k_0} sm_s{IM_I}\rangle.
 \label{eq:nad4c}
 \end{eqnarray}
 If the momentum ${\bf k_0}$ is chosen to be along the $z$-axis, the orbital
 angular momentum $l$ has no projection along this direction. Since we are 
 dealing with axially symmetric potentials, we set the azimuthal angle $\phi=0$.
 The scattering amplitude thus takes the form
    \begin{eqnarray}
  f_{I'M_I'sm_s':IM_Ism_s}(E,\theta)
 &=&\frac{4\pi}{k_0}\sum\limits_{JM} \sum\limits_{ll'j_pj_p'}
C(Ij_pJ,M_I M-M_I M)\;C(lsj_p,M-M_I-m_s m_s M-M_I)\cr\cr
 &\times&\;C(l'sj_p,M-M_I'-m_s' m_s' M-M_I')\;C(I'j_p'J,M_I' M-M_I' M)\cr\cr
 &\times& \kappa_{\alpha\alpha_0}^{J}(E)\;Y_{l'M-M_I'-m_s'}(\theta,0) Y_{l0}^*({0,0}).
 \label{eq:nad4d}
 \end{eqnarray}
  Finally, to obtain the differential cross section we take the square of the
  scattering amplitude $|f_{I'M_I'sm_s':IM_Ism_s}(E,\theta)|^2$. If the spin
  projections are not measured, the cross section is evaluated by averaging
  over the initial magnetic quantum numbers and summing over the final ones.
  This leads to the unpolarized differential cross section
   \begin{equation}
   \frac{d\sigma(\theta)}{d\Omega}= \frac{1}{(2I+1)(2s+1)}\sum\limits_{M_{I}'m_{s'}}\sum\limits_{M_Im_s}
    \Big |f_{I'M_{I}'sm_{s'}:IM_Ism_s}(E,\theta) \Big|^2.
    \label{eq:nad4e}
   \end{equation}

\subsection{Protons}
\label{appx-pascat}

 The coupled-channel calculations for proton-nucleus
 systems proceed similarly to those of the neutron-nucleus
 systems. However, due to the presence of the Coulomb force
 the point-Coulomb is separated via the Gell-Mann-Golberger
 relation~\cite{RodbergThaler} and the nuclear amplitude
 is calculated in the basis of Coulomb scattering states.
 The full proton-nucleus scattering amplitude thus takes the
 form~\cite{ThompsonNunes}
     \begin{eqnarray}
  f_{I'M_I'sm_s':IM_Ism_s}(E,\theta)
 &=&f_{c}(\theta)\delta_{I'M_I'sm_s':IM_Ism_s}\cr\cr 
 &+& \frac{4\pi}{k_0}\sum\limits_{JM} \sum\limits_{ll'j_pj_p'}e^{i[\sigma_{l'}(k_\alpha)+\sigma_l(k_0)]}
C(Ij_pJ,M_I M-M_I M)\;C(lsj_p,M-M_I-m_s m_s M-M_I)\cr\cr
 &\times&\;C(l'sj_p,M-M_I'-m_s' m_s' M-M_I')\;C(I'j_p'J,M_I' M-M_I' M)\cr\cr
 &\times& \kappa_{\alpha\alpha_0}^{cJ}(E)\;Y_{l'M-M_I'-m_s'}(\theta,0) Y_{l0}^*({0,0}).
 \label{eq:pad4d}
 \end{eqnarray}
 Here $f_{c}(\theta)$ is the Rutherford amplitude and $\sigma_{l'}(k_0)$ the Coulomb phase shift. The dimensionless $\kappa_{\alpha\alpha_0}^{cJ}(E)$ is related to the Coulomb-distorted multichannel $t$ matrix
 $T_{\alpha\alpha_0}^{cJ}(k^\alpha_0,k^{\alpha_0}_0;E)$ by 
 Eq.~(\ref{eq:nad3d}). The unpolarized differential cross sections
 computed are according to Eq.~(\ref{eq:nad4e}).  
 
 %


\newpage
\section{ Numerical Aspects}
\label{appx-B}

\subsection{Neutron Optical Potentials}
\label{appx-na}

To evaluate the separable $t$ matrix
of Eq.~(\ref{eq:mch21c}), the multichannel
half-shell $t$ matrices $T_{\alpha\rho}^J(E_i)\big| k_i^\rho\big\rangle$ and 
coupling matrix $\tau_{ij}
  ^{\rho\sigma}(E)$ are required
as input. The half-shell $t$ matrices
are obtained as solutions of Eq.~(\ref{eq:multi3g}).
The matrix $\tau_{ij}
  ^{\rho\sigma}(E)$ is determined by solving Eq.~(\ref{eq:mch24a}), where
the matrix $ R^{\rho\sigma}_{ij}(E)$ is calculated
according to Eq.~(\ref{eq:mch24b}). For the purpose
of evaluating $\tau_{ij}
  ^{\rho\sigma}(E)$ numerically, a different notation for the
channel and energy indices is adopted. Each combination  $\{i,\rho\}$ is denoted by a single index $a$. We proceed by expressing Eqs.~(\ref{eq:mch24a}) - (\ref{eq:mch24b}) in the form
\begin{eqnarray}
  \mathcal{M}_{ab}(E)&\equiv& [R(E)\cdot\tau(E)]_{ab},
 \label{eq:esep10a}
 \end{eqnarray}
  where
 \begin{eqnarray}
  R_{ab}(E)&=& \langle k_a\;a JM\big| T(E_a) +T(E_a)G_0(E_b)T(E_b)\big|
  b JM\; k_b\big\rangle\cr\cr  &-&
\sum\limits_{c}\mathcal{M}_{ac}^{}(E)\langle k_c\; c JM\big| T(E_c)G_0(E)T(E_b)
\big| b JM\; k_b\big\rangle,
  \label{eq:esep10}
 \end{eqnarray}
with
  \begin{eqnarray}
  \mathcal{M}_{ac}^{}(E)&\equiv&\left[\mathcal{U}^{e}(E){\cal U}^{-1}\right]_{ac}.
  \label{eq:mch24c}
  \end{eqnarray}
The explicit momentum space expression for $R(E)$ is given as
\begin{eqnarray}
 R_{ab}(E)&=& {T}^{J}_{\alpha_a\alpha_b}(k_b,k_a;E_a) +
 \sum\limits_{\beta}\int\limits_0^\infty dp\;p^2\; {T}^{J}_{\alpha_a,\beta}(p,k_a;E_a)
 G_{0\beta}(E_b)T^{J}_{\beta\alpha_b}(p,k_b;E_b) \cr\cr&-&\sum\limits_{c}\mathcal{M}_{ac}(E)
 \sum\limits_\beta\int\limits_0^\infty dp\;p^2\;
  {T}^{J}_{\alpha_c\beta}(p,k_c;E_c)G_{0\beta}(E)T^J_{\beta\alpha_b}(p,k_b;E_b).
  \label{eq:esep11}
 \end{eqnarray} 
  The coupling matrix $\tau(E)$ is thus determined from Eq.~(\ref{eq:esep10a}) 
  so that the separable multichannel $t$ matrix is given as
 \begin{eqnarray}
  t_{\alpha\beta}^J(k',k;E)&=&\sum\limits_{\rho\sigma}\sum\limits_{ij} T^{J}_{\alpha\rho}(k',k_i^\rho;E_i)
  \tau_{ij} ^{\rho\sigma}(E) {T}^{J}_{\sigma \beta }(k,k_j^\sigma;E_j),\cr\cr
  &=&\sum\limits_{ab} T^{J}_{\alpha a}(k',k_a;E_a)\tau_{ab}(E)
   {T}^{J}_{b \beta}(k,k_b;E_b).
 \label{eq:esep9}
 \end{eqnarray}
 If the potential $U$ is energy-independent, the matrix
 $\mathcal{M}_{ac}(E)$ reduces to an identity matrix. 
 Consequently, the eEST scheme reduces to an energy-independent separable representation. In this limit 
 the matrix $R_{ab}(E)$ is given by
   \begin{eqnarray}
  R_{ab}(E)&=& {T}^{J}_{\alpha_a\alpha_b}(k_b,k_a;E_a) \cr\cr
  &&+
 \sum\limits_{\beta}\int\limits_0^\infty dp\;p^2\; {T}^{J}_{\alpha_a,\beta}(p,k_a;E_a)
 G_{0\beta}(E_b)T^{J}_{\beta\alpha_b}(p,k_b;E_b) \cr\cr
 &&-\sum\limits_{\beta}\int\limits_0^\infty dp\;p^2\;
  {T}^{J}_{\alpha_a\beta}(p,k_a;E_a)G_{0\beta}(E)T^J_{\beta\alpha_b}(p,k_b;E_b).
  \label{eq:sepc5}
 \end{eqnarray} 
\subsection{Proton-nucleus Optical Potentials}
\label{appx-pa}

We note that the separable $t$ matrix given
by Eqs.~(\ref{eq:cmch21c}) - (\ref{eq:cmch24b}) has a similar form
 as the one obtained for neutron optical potentials, except that
\begin{enumerate}[(a)]

\item  the half-shell $t$ matrices $T_{\alpha\rho}^J(E_i)\big| k_i^\rho\big\rangle$
are replaced by the Coulomb-distorted half-shell
$t$ matrices $T_{\alpha\rho}^{cJ}(E_i)\big| k_i^\rho\big\rangle$,

\item the Coulomb propagator $G_C(E)=[E-H_0-V^C+i\epsilon]^{-1}$ replaces
 the free propagator $G_0(E)$.
   
\end{enumerate}
The Coulomb-distorted $t$ matrix elements fulfill a set of LS equations
 \begin{equation}
  T_{\alpha\alpha_0}^{cJ}(k',k;E)=U_{\alpha\alpha_0}^{cJ}(k',k)
  +\sum\limits_{\alpha'}\int\limits_{0}^\infty dp p^2\;U_{\alpha\alpha'}^{cJ}(k,p)G_{C\alpha'}(E,p)
  T_{\alpha'\alpha_0}^{cJ}(p,k;E).
  \label{eq:cmulti3g}
  \end{equation}
 Evaluating the Coulomb propagator $G_C(E)$ in the
 Coulomb basis leads to
 \begin{eqnarray}
 G_{C\alpha}(E,p)=G_{0\alpha}(E,p)=
 \left(E-\varepsilon_{\alpha}-p^2/2\mu_{\alpha}+i\epsilon\right)^{-1},
 \label{eq:cmulti4a}
 \end{eqnarray}
 where  $\mu_\alpha$ is the reduced mass in channel $\alpha$.
Here $U_{\alpha\alpha_0}^{cJ}(k',k)$  are the potential matrix elements
in the Coulomb basis. A direct evaluation of $U_{\alpha\alpha_0}^{cJ}(k',k)$ in momentum space
is extremely difficult since the Coulomb wavefunctions
are singular. Instead, it is evaluated using the non-singular coordinate space Coulomb wavefunctions as described in Ref.~\cite{Elster:1993dv}.

 Repeating the steps outlined in Section~\ref{appx-na} we obtain
 \begin{eqnarray}
  t_{\alpha\beta}^{cJ}(k',k;E)&=&\sum\limits_{\rho\sigma}\sum\limits_{ij} {T^c}^{J}_{\alpha\rho}(k',k_i^\rho;E_i)
  \tau_{ij} ^{c,\;\rho\sigma}(E) {T^c}^{J}_{\sigma \beta }(k,k_j^\sigma;E_j),\cr\cr
  &=&\sum\limits_{ab} {T^c}^{J}_{\alpha a}(k',k_a;E_a)\tau_{ab}^c(E)
   {T^c}^{J}_{b \beta}(k,k_b;E_b).
 \label{eq:cesep9}
 \end{eqnarray}
 The matrix $R_{ab}^c(E)$ fulfills
\begin{eqnarray}
  \mathcal{M}_{ab}^c(E)&\equiv& [R^c(E)\cdot\tau^c(E)]_{ab},
 \label{eq:cesep10a}
 \end{eqnarray}
where
\begin{eqnarray}
 R_{ab}^c(E)&=& {T^c}^{J}_{\alpha_a\alpha_b}(k_b,k_a;E_a) +
 \sum\limits_{\beta}\int\limits_0^\infty dp\;p^2\; {T^c}^{J}_{\alpha_a,\beta}(p,k_a;E_a)
 G_{C\beta}(E_b){T^c}^{J}_{\beta\alpha_b}(p,k_b;E_b) \cr\cr&-&\sum\limits_{a}\mathcal{M}_{ad}^{c}(E)
 \sum\limits_\beta\int\limits_0^\infty dp\;p^2\;
  {T^c}^{J}_{\alpha_d\beta}(p,k_d;E_d)G_{C\beta}(E){T^c}^J_{\beta\alpha_b}(p,k_b;E_b).
  \label{eq:cesep11}
 \end{eqnarray} 
The matrix ${M}_{ad}^{c}(E)$ is given by
  \begin{eqnarray}
  \mathcal{M}_{ad}^{c}(E)&\equiv&\left[\mathcal{U}^{ce}(E)({\cal U}^c)^{-1}\right]_{ad}.
  \label{eq:cmch24c}
  \end{eqnarray}
%

\newpage
\section{Deformed Optical Model Potentials}
\label{domps}

Spherical optical model potentials (OMPs) are usually based on Woods-Saxon functions and depend on the distance between
the nucleon and the surface of the nucleus, $r-R$. Nuclear deformations naturally lead to a deformed Optical Model Potential (DOMP).
The evaluation of the DOMP presented here follows closely that of Refs.~\cite{Chase1958,Buck1963}.
For a DOMP the Woods-Saxon functions depend on the orientation
of the nucleus so that $f_{ws}(r,\theta,a,R)\equiv f_{ws}(\tilde r(\theta), a,R)$
where the shifted radius is given as
 \begin{equation}
  \tilde{r}(\theta)=r-\delta(\mathbf{\hat\xi}\cdot\mathbf{\hat r}).
  \label{eq:multi1}
 \end{equation}
Here $\mathbf{\hat\xi}$ represents the orientation of the 
nucleus relative to a space fixed coordinate frame.
The shift function $\delta(\mathbf{\xi}\cdot\mathbf{r})$  is expanded in
multipoles
 \begin{eqnarray}
  \delta(\mathbf{\hat\xi}\cdot\mathbf{\hat r})&=&\sum\limits_{\lambda\ne 0}\delta_\lambda Y_\lambda^{0}(\theta,0),\cr\cr
  &=& \sum\limits_{\lambda\ne0,\; \mu}\frac{\sqrt{4\pi}}{\sqrt{2\lambda+1}}\delta_\lambda 
  {Y_\lambda^{\mu}}^*(\mathbf{{\hat\xi}})Y_\lambda^{\mu}(\mathbf{{\hat r}}). 
  \label{eq:rp2}
 \end{eqnarray}
 The angle $\theta$ is defined by $\cos\theta\;=\mathbf{\hat\xi}\cdot\mathbf{\hat r}$. In the reference frame rotating with the nucleus, 
 $\theta$ is simply the zenith angle of the relative vector $\mathbf{r}$. The deformation
 length $\delta_\lambda$ is proportional the maximum shift in the radius $r$
 at each multipole. 
 Each term of the OMP is deformed independently, and has the multipole
 expansion 
  \begin{equation}
  \hat U(\mathbf{\xi},\mathbf{r})=
  \sum\limits_{\lambda\mu} \sqrt{4\pi}\hat U_\lambda(r) D^\lambda_{\mu,0}(\mathbf{{\hat\xi}})Y_\lambda^{\mu}
  (\mathbf{{\hat r}}),
  \label{eq:multi2}
 \end{equation}
 where the rotational matrix is given by
 \begin{equation}
  D^\lambda_{\mu,0}(\mathbf{{\hat\xi}})=\frac{\sqrt{4\pi}}{\sqrt{2\lambda+1}} {Y_\lambda^{\mu}}^*(\mathbf{{\hat\xi}}).
  \label{eq:multi3}
 \end{equation}
The expansion in Eq.~(\ref{eq:multi2}) can be written as
   \begin{eqnarray}
  \hat U(\mathbf{\xi},\mathbf{r})&=&
  \sum\limits_{\lambda\mu} \sqrt{4\pi}\hat U_\lambda(r)\frac{\sqrt{4\pi}}{\sqrt{2\lambda+1}} {Y_\lambda^{\mu}}^*
  (\mathbf{{\hat\xi}})Y_\lambda^{\mu}(\mathbf{{\hat r}})\\
  &=&   \sum\limits_{\lambda} \sqrt{4\pi} \hat U_\lambda(r)Y_\lambda^{0}(\theta,0),\cr
      &=&  {4\pi}\sum\limits_{\lambda}(-1)^\lambda \hat U_\lambda(r) [Y_\lambda(\mathbf{{\hat\xi}})
  \times Y_\lambda(\mathbf{{\hat r}})]_0^0.
  \label{eq:multi4}
 \end{eqnarray}
 For each multipole $\lambda$, the potential is given by the angular integral
 \begin{eqnarray}
  \hat U_\lambda(r)&=& \sqrt{\pi}\int\limits_{-1}^{1}d\cos\theta\; U(\tilde r)Y_\lambda^{0}(\theta,0),\cr
   &=& \frac{1}{2}\sqrt{2\lambda+1}\int\limits_{-1}^{1}d\cos\theta\; U(\tilde r)P_\lambda(\cos\theta).
   \label{eq:multi5}
 \end{eqnarray}
 This integral is zero for odd $\lambda$, which implies that $\hat U$ can only change the 
 spin of the nucleus by an even number. For small deformations we can perform a Taylor expansion around
 $\tilde r =r$
 \begin{equation}
   U(\tilde r)= U(r)- U'(r) \sum\limits_{\lambda\ne 0}\delta_\lambda Y_\lambda^{0}(\theta,0)+...\; .
  \label{eq:multi6}
 \end{equation}
The monopole term is given by
 \begin{eqnarray}
  \hat U_0&=& U(r),
  \label{eq:multi7}
 \end{eqnarray}
 and for $\lambda>0$ one obtains
 \begin{equation}
 \hat U_\lambda= -\frac{1}{\sqrt{4\pi}}\delta_\lambda U'(r). 
  \label{eq:multi8}
 \end{equation}
 An alternative way of representing the deformation involves defining
 the deformation parameter $\beta_\lambda=\delta_\lambda/R$. It
 is a measure of the deformation relative to the radius of the 
 spherical potential. 
 
 A parametrization of a DOMP must specify, in addition
 to the the optical potential shape $U(r)$, the deformation length
 $\delta_\lambda$ or the deformation parameter $\beta_\lambda$.
 The multipole potential can then be evaluated
 according to Eq.~(\ref{eq:multi5}). For  small deformation lengths
 the multipole potentials are evaluated using Eqs.~(\ref{eq:multi7}-\ref{eq:multi8}).
 In practice, only a few terms of the multipole expansion are necessary
 to describe elastic and inelastic nucleon-nucleus scattering. In order to constrain
 the DOMP, both elastic and inelastic scattering data are necessary.

\subsection{The Olsson 89 Deformed  Optical Model Potential}
\label{olsson}

The Olsson~89~\cite{Olsson:1989npa} DOMP has the form
\begin{eqnarray}
 -U(r,E)&=&V_r(E)\;f_{ws}(r,a_r,R_r) 
 +2V_{so}(E)\left(\frac{-1}{r}\right)\frac{d}{dr}f_{ws}(r,a_{so},R_{so})\mathbf {\;l\cdot\boldsymbol\sigma}\cr
 &+&iW_s(E)(-4a_s)\frac{d}{dr}f_{ws}(r,a_s,R_s).     
 \label{eq:opt1}
\end{eqnarray}
The real and imaginary strengths are indicated by $V(E)$ and $W(E)$. The indices $r$, $so$, $i$, and $s$ denote the
real volume, spin-orbit, imaginary volume, and surface potential terms respectively. The imaginary surface
term  is included to simulate the effects of a surface-peaked absorption at low energies, while at higher
energies volume absorption dominates.  
 The Woods-Saxon (WS) function is given by
\begin{eqnarray}
f_{ws}(r,a,R)=\left[1+\exp\left(\frac{r-R}{a}\right)\right]^{-1},
\label{eq:opt0o}
\end{eqnarray}
where  $a$ and $R$ are the diffusiveness and radius.

 The Olsson 89~\cite{Olsson:1989npa}  DOMP is fitted to neutron elastic and inelastic
 scattering cross sections from $^{12}$C. It is appropriate for the incident
 neutron energies starting from 16 to 26~MeV. Only the real volume, imaginary surface,
 and real spin-orbit terms have been included. The multipole expansion is truncated at $\lambda= 4$ 
 with the deformation parameters given as
  \begin{equation}
  \beta_2=-0.65 \text{~and~} \beta_4= 0.05. 
  \label{eq:multi9}
  \end{equation} 
  For the readers convenience the remaining parameters are shown in Table~\ref{table:multi1}.
  \begin{table}[ht!]
  \centering
  \begin{tabular}{cccc}
   \hline\hline
   \\
    Strength [MeV] & Radius [fm] & Diffusiveness [fm]\\
    \\
    \hline
    \\
     $V_r=64.02-0.674E_n$ &  $R_r=1.093A^{1/3}$&  $a_r=0.619$\\
     \\
     $W_s=1.16+0.251E_n$& $R_s=1.319A^{1/3}$& $a_s=0.327$\\
     \\
     $V_{so}= 6.2$& $R_{so}=1.050A^{1/3}$& $a_{so}=0.550$\\
     \\
  \hline
  \end{tabular}
  \vspace{3mm}
    \caption{Deformed optical potential parameters adjusted to n+$^{12}$C elastic and inelastic scattering
           data. These parameters are taken from Ref.~\cite{Olsson:1989npa}. Here $E_n$ is the neutron energy
           in the laboratory frame and should be given in MeV.
 }
  \label{table:multi1}
  \end{table}

\subsection{The Meigooni~85 Deformed Optical Model Potential} 
\label{meigooni84}

The Meigooni~85~\cite{Meigooni:1985iwq} DOMP has the form
\begin{eqnarray}
 -U(r,E,A)&=&V_r(E,A)\;f_{ws}(r,a_r,R_r)\cr 
 &+&2\Big(V_{so}(E,A)\Big)\left(\frac{-1}{r}\right)\frac{d}{dr}f_{ws}(r,a_{so},R_{so})\mathbf {\;l\cdot\boldsymbol\sigma}\cr
 &+&i\left[ W_i(E,A)\;f_{ws}(r,a_i,R_i)
 +W_s(E,A)(-4a_s)\frac{d}{dr}f_{ws}(r,a_s,R_s)\right],     
 \label{eq:opt1m}
\end{eqnarray}
where the Woods-Saxon function is given by Eq.~(\ref{eq:opt0o}).
 The parameters are determined from neutron scattering data. 
The proton OMP is obtained by simply adding the short-ranged Coulomb term to the neutron OMP.  The deformation parameters have the values
 $\beta_2=$-0.61 and $\beta_4=$0.05.  
 The remaining parameters are shown in Table~\ref{table:multi2}  for the readers convenience.
  \begin{table}[ht!]
  \centering
  \begin{tabular}{cccc}
   \hline\hline
   \\
    Strength [MeV] & Radius [fm] & Diffusiveness [fm]\\
    \\
    \hline
    \\
     $V_r=64.02-0.674E_n$ &  $R_r=1.093A^{1/3}$&  $a_r=0.619$\\
     \\
     \hline
     \\
     $W_i(E)=0$ for $E\le 20$& $R_i=1.22A^{1/3}$& $a_i=0.478+0.0043E$\\
     \\
     $W_i(E)=15.5\left[1-2/(1+\exp(E-20)/25)\right]$ for $E>20$&&\\
     \\ 
     \hline
     \\
 $W_s(E)= e^{0.095E}$ for $E\le 21$~MeV&&\\
 \\
 $W_s(E)= 10.29-0.145E$ for $21 \le E < 71$~MeV & $R_s=1.25A^{1/3}$ & $a_s=0.27$\\
 \\
 $W_s(E)= 0$ for $E> 71$~MeV&&\\   
 \\   
 \hline
 \\
 $V_{so}= 6.2$& $R_{so}=1.050A^{1/3}$& $a_{so}=0.550$\\
 \\
  \hline
  \end{tabular}
  \vspace{3mm}
    \caption{Parameters for the Meigooni~85~\cite{Meigooni:1985iwq} DOMP. Here $E$ is the neutron energy
           in the laboratory frame and should be given in MeV.
 }
  \label{table:multi2}
  \end{table}
 
\end{document}